\documentclass[12pt,draftclsnofoot,onecolumn]{IEEEtran}

\ifCLASSINFOpdf
  
\else
  
\fi

\usepackage{setspace}
\usepackage[ruled,vlined]{algorithm2e}
\usepackage{cite}
\usepackage{graphicx}
\usepackage{subfig}
\usepackage{amsfonts}
\usepackage{amsmath}
\usepackage{amsthm}
\usepackage{mathdots}
\usepackage{bm}

\begin{document}

\begin{spacing}{1.25}

\title{Joint Location and Communication Study for Intelligent Reflecting Surface Aided Wireless Communication System}

\author{
        Rui Wang,
        Zhe Xing,
        and Erwu Liu% <-this % stops a space
\thanks{The authors are with the College of Electronics and Information Engineering, Tongji University, Shanghai 201804, China. Rui Wang is also with the Shanghai Institute of Intelligent Science and Technology, Tongji University, Shanghai, China. (e-mails: zxing@tongji.edu.cn; ruiwang@tongji.edu.cn;  erwuliu@tongji.edu.cn).
}}% <-this % stops a space
%\thanks{\textit{Corresponding Author: Rui Wang}.}}% <-this % stops a space

\markboth{}%
{\MakeLowercase{\textit{et al.}}}

\maketitle

\begin{abstract}

\begin{spacing}{1.3}

Intelligent reflecting surface (IRS) is a novel burgeoning concept, which possesses advantages in enhancing wireless communication and user localization, while maintaining low hardware cost and energy consumption. Herein, we establish an IRS-aided mmWave-MIMO based joint localization and communication system (IMM-JLCS), and probe into its performance evaluation and optimization design. Specifically, first, we provide the signal, channel and estimation error models, and contrive the working process of the IMM-JLCS in detail. Then, by configuring appropriate IRS phase shifts, we derive the closed-form expressions of the Cramér-Rao Lower Bound (CRLB) of the position/orientation estimation errors and the effective achievable data rate (EADR), with respect to the time allocation ratio of the beam alignment and localization stage (BALS). Subsequently, we investigate the trade-off between the two performance metrics, for which we propose a joint optimization algorithm. Finally, we carry out simulations and comparisons to view the trade-off and validate the effectiveness of the proposed algorithm, in the presence of distinct levels of estimation uncertainty and user mobility. Our results demonstrate that the proposed algorithm can find the joint optimal solution for the position/orientation estimation accuracy and EADR, with its optimization performance being robust to slight localization or channel estimation errors and user mobility.

\end{spacing}

\end{abstract}


\IEEEpeerreviewmaketitle

\newpage

\section{Introduction}
\IEEEPARstart{T}{he} fifth-generation (5G) mobile communication network has been standardized and commercially deployed in part since the first quarter of 2020, while the researches on the sixth-generation (6G) mobile communication have already begun to advance \cite{6G2,6G3,6G1}. With an enormous amount of worldwide mobile communication devices to be served, various key enabling technologies, including the millimeter-wave (mmWave), massive multiple-input-multiple-output (MIMO) and ultra-dense network (UDN), have been developed to fulfil the requirements of Gbps level of achievable data rate, high spectral efficiency, mass connectivity, ultra-reliability and low latency \cite{6G1}. 
While these technologies possess prominent advantages in improving the wireless communication performance, they are still facing several challenging and inevitable issues. First, the mmWave is susceptible to blockage and suffers from serious free-space propagation loss in the atmosphere due to its high frequency \cite{mmWave}. Second, the massive MIMO and UDN consist of large-scale antenna arrays and serried base stations (BSs), resulting in high hardware cost and energy consumption \cite{MIMO,UDN}. In view of these problems, the future 6G will focus more on the exploration of novel communication paradigms on the foundation of the current 5G.

Recently, the prospective alteration of the communication paradigm is enabled by a novel burgeoning concept, named Intelligent Reflecting Surface (IRS), or Reconfigurable Intelligent Surface (RIS), Large Intelligent Surface (LIS), which is proposed by the inspiration of the idea of manipulating the wireless communication environment \cite{Propose IRS 1, Propose IRS 2}. The IRS is a two-dimensional (2D) planar reflection array, composed of a large quantity of low-cost passive reflecting units, which can induce reconfigurable phase shifts on the impinging signal waves before reflecting them to the receiving terminals \cite{IRS Survey}. As it can usually be fabricated with cheap positive intrinsic-negative (PIN) diodes \cite{L.Dai2020(Access)} or varactor diodes \cite{J.Y.Lau2012(TAP)}, and be deployed almost anywhere to establish a strong virtual line-of-sight (VLoS) link without the necessity of power-consuming radio-frequency (RF) chains \cite{Q.Wu2020(CM)}, it is envisioned as a promising hardware solution to the problems of the propagation limit, hardware cost and energy consumption.
Up to now, the IRS has been listed in "White Paper
on Broadband Connectivity in 6G" \cite{6G White Paper} as a candidate technology in the future 6G mobile communication network, and has been extensively adopted in various communication scenarios to enhance the wireless data transmission, e.g. to improve the spectral and energy efficiency \cite{Spectral Efficiency 1,Spectral Efficiency 2,Spectral Efficiency 3,Spectral Efficiency 4}, maximize the achievable data rate \cite{ACR Maximization 1,ACR Maximization 2}, achieve the secure wireless transmission \cite{Secure Transmission 2, Secure Transmission 4}, design the index-modulation scheme \cite{Index Modulation 1,Index Modulation 2}, transfer passive information \cite{Passive Information 1,Passive Information 2}, \textit{et al}., and been investigated in terms of the channel capacity \cite{Channel Capacity}, outage probability \cite{Outage Probability}, coverage \cite{Coverage}, hardware impairments \cite{Hardware Impairments 1}, \textit{et al}., of the IRS-aided wireless communication system. 

In addition to improving the communication performance, assisting the user localization is also an important potential functionality of the IRS to be excavated. It is noted that the mmWave and massive MIMO can be amalgamated to localize the mobile user (MU) based on the channel parameters (e.g. angle of arrival/departure (AOA/AOD), time delay, \textit{et al}.) \cite{mmWave MIMO Localization}, owing to the “quasi-optical” propagation property of the mmWave signals \cite{Quasi-optical} and the compact directional and steerable large antenna arrays of the massive MIMO \cite{MIMO}. Among the previous studies on the mmWave-MIMO based positioning systems \cite{Shahmansoori2018(TWC), Ghaseminajm2020(TVT), Wang2019(TSP)}, investigating the Cramér-Rao Lower Bound (CRLB) of the position and orientation estimation errors in the presence of scatterers \cite{Shahmansoori2018(TWC)}, I/Q imbalance \cite{Ghaseminajm2020(TVT)}, multipath fading \cite{Wang2019(TSP)}, \textit{et al}., and designing effective estimation algorithms based on compressed sensing (CS) \cite{Shahmansoori2018(TWC)}, maximum-likelihood (ML) \cite{Wang2019(TSP)}, \textit{et al}., are two of the most typical research directions followed with interest. Because of the mmWave's susceptibility to blockage, some researchers have already begun to explore the application potential of the IRS in the mmWave-MIMO based localization system \cite{S.Hu2018(TSP),J.He2020(VTC),J.He2020(WCNC), A.Elzanaty2020(ArXiv), H.Zhang2021(CL), H.Zhang2020(ArXiv), X.Hu2020(TCOM)}. As an early research, S. Hu, \textit{et al}. \cite{S.Hu2018(TSP)}, first introduced the IRS to the wireless localization system and derived the CRLB of the localization error. Afterwards, J. He, \textit{et al}. \cite{J.He2020(VTC),J.He2020(WCNC)}, leveraged the IRS to assist the positioning in a 2D mmWave localization system, and testified its capability of improving the localization performance. By considering a more practical system model, A. Elzanaty, \textit{et al}. \cite{A.Elzanaty2020(ArXiv)}, investigated the similar problem in the 3D environment, making the analytical results conform to the real-world scenario; H. Zhang, \textit{et al}. \cite{H.Zhang2021(CL), H.Zhang2020(ArXiv)}, localized the MUs based on the received signal strength (RSS) in an indoor environment, and utilized the IRS to improve the differences of the RSS between adjacent location blocks. In turn, X. Hu, \textit{et al}. \cite{X.Hu2020(TCOM)}, adopted the user's location information, provided by the global positioning system (GPS), to design the IRS phase shifting matrix.

Nevertheless, the aforementioned works still left a few research gaps to be filled:
First, the IRS-aided mmWave-MIMO based joint localization and communication scheme was not considered, which, however, would gradually become popularized and universal in the future mobile communication network. It is remarkable that when the communication and localization approaches are integrated in one system, a trade-off exists between the positioning accuracy and effective achievable data rate (EADR) \cite{Destino2017(ICC), Kumar2018(ICL), Destino2018(WCNC), Koirala2019(Access), Ghatak2018(VTC)}. From this perspective, G. Destino, \textit{et al}. \cite{Destino2017(ICC), Kumar2018(ICL), Destino2018(WCNC)}, performed some important works by dividing a fixed communication duration into two separate time slots for localization and effective data transmission, respectively, and inquiring into the trade-off between the positioning quality and EADR. 
R. Koirala, \textit{et al}. \cite{Koirala2019(Access)}, also studied the trade-off from the perspective of the time allocation, and formulated optimization problems to optimize the localization and EADR performances.
G. Ghatak, \textit{et al}. \cite{Ghatak2018(VTC)}, derived the CRLB for the estimation of the distance between a mobile user and its serving BS, and investigated the trade-off by allocating the total transmit power for the positioning and effective data transmission. However, in these researches, only BSs and MUs were taken into account, with the LoS link assumed to be available in between. If the LoS link is obstructed, it is necessary to introduce the IRS into the mmWave-MIMO based joint localization and communication system, in order to maintain or improve both the localization and communication performances. Besides, the IRS configuration (e.g. the number of the reflecting elements, the phase shifts) may influence the trade-off, which deserves to be investigated in depth as well.
Second, according to the trade-off between the positioning accuracy and EADR, with distinct system settings, the two performance metrics may not simultaneously reach their own maximums, but instead, can possibly achieve their joint optimal point. Therefore, a joint optimization algorithm is required for guiding the optimal system setup. Third, the IRS phase shifts need to be adjusted to cater for the localization and communication requirements, but the phase shift adjustment depends on the MU's position information in turn. Thus, a specific framework of the system's working process should be designed to facilitate the IRS configuration in the real-world application scenarios.

To the best of our knowledge, we have not found the related works carried out by considering the above three aspects. Consequently, in this article, we first establish an IRS-aided mmWave-MIMO based joint localization and communication system (IMM-JLCS) and design a framework of its working process, and then probe into the trade-off and joint optimization on the positioning accuracy and EADR, with our contributions summarized as follows.

%For instance, A. Shahmansoori, \textit{et al}. \cite{Shahmansoori2018(TWC)}, derived the CRLB on the estimation errors of the user's position and rotation angle in the presence of scatterers, and designed a three-stage estimation algorithm. 
%based on the distributed compressed sensing-simultaneous orthogonal matching pursuit (DCS-SOMP), the space-alternating generalized expectation maximization (SAGE) and the extended invariance principle (EXIP).
%Z. A. Shaban, \textit{et al}. \cite{Shaban2018(TWC)}, analysed the position and orientation error bounds for the general uplink and downlink localization in the three-dimensional (3D) communication scenario.
%F. Ghaseminajm, \textit{et al}. \cite{Ghaseminajm2020(TVT)}, studied the impact of the I/Q imbalance on the position and orientation error bounds when using the quadrature amplitude modulation (QAM) in 5G mmWave communication systems.
%Y. Wang, \textit{et al}. \cite{Wang2019(TSP)}, evaluated the effect of the multipath propagation on the localization performance, 
%and proved the feasibility of the joint spatiotemporal multipath mitigation with general non-orthogonal waveforms when utilizing the large-scale antenna arrays. 
%and developed a maximum-likelihood based localization method. 

\begin{itemize}
\item[•] We construct a 3D IMM-JLCS in the presence of an obstacle which blocks the LoS link. In this IMM-JLCS, first, we divide a communication period with a fixed duration into three stages, i.e. beam alignment and localization stage (BALS), effective data transmission stage (EDTS) and joint optimization stage (JOS), for position/orientation estimation, effective data transmission, and joint optimization on the localization and communication performances, respectively. Then, we design a complete framework of the working process for the considered system within each communication period. 

\item[•] We calculate the CRLBs of the position/orientation estimation errors and the EADR of the effective data transmission to evaluate the potential localization and communication performances, and derive their approximate closed-form expressions with respect to the time allocation ratio of the BALS by configuring appropriate IRS phase shifts. 

\item[•] Under different time allocation ratio, we investigate the trade-off between the positioning accuracy and EADR.
Based on the insight provided by the trade-off, we formulate a joint optimization problem to optimize the time allocation ratio, in order to find a joint optimal solution of the CRLB and EADR. By solving this problem with the Lagrangian multipliers and the Karush-Kuhn-Tucker (KKT) conditions, we finally propose a joint optimization algorithm for the two performance metrics.

\item[•] In order to view the trade-off and validate the effectiveness and robustness of the proposed algorithm, we carry out simulations in the presence of distinct levels of 1) user mobility and 2) channel and position/orientation estimation errors. Moreover, we numerically compare the designed IRS phase shifts with random IRS phase shifts in terms of the potential localization and communication performances, for the purpose of evaluating the performance improvement brought by the IRS phase shift configuration in our work.
\end{itemize}

The remainder of this article is organized as follows. In Section II, we present the system model and the working process of the IMM-JLCS. In Section III, we calculate the position/rotation error bounds and the EADR, and derive their approximate closed-form expressions in relation to the time allocation ratio. In Section IV, we discuss the trade-off between the two performance metrics, and propose the joint optimization algorithm. In Section V, we carry out simulations to view the numerical results and make performance comparisons. In Section VI, we draw the overall conclusions.

\textit{Notations:} Boldfaces and italics stand for the vectors or matrices and the constants or variables, respectively. $\mathbf{X}^T$, $\mathbf{X}^*$, $\mathbf{X}^H$ and $\mathbf{X}^{-1}$ represent the transpose, conjugate, conjugate-transpose and inverse of $\mathbf{X}$. $\left[\mathbf{X}\right]_{(a,b)}$ represents the $(a,b)$-th element in $\mathbf{X}$. $tr(\mathbf{X})$ denotes the trace of $\mathbf{X}$. $diag(x_1,x_2,...,x_n)$ stands for a diagonal matrix with its diagonal elements of $(x_1,x_2,...,x_n)$. $\|.\|$ and $\|.\|_2$ symbolize the $\ell_1$-norm and $\ell_2$-norm. $\otimes$ and $\odot$ symbolize the Kronecker product and Hadamard product. $\mathfrak{Re}\{x\}$ and $\mathfrak{Im}\{x\}$ are the real part and imaginary part of $x$. $\widehat{x}$ denotes the estimate of $x$. $\mathbb{E}_{\mathbf{a}}[\mathbf{X}]$ denotes the expectation of $\mathbf{X}$ on $\mathbf{a}$ if $\mathbf{X}$ is a random matrix in relation to $\mathbf{a}$. $\bigtriangledown_{\mathbf{a}}\mathbf{X}$ represents the gradient of $\mathbf{X}$ with respect to $\mathbf{a}$, while $\frac{\partial x}{\partial a}$ represents the partial derivative of $x$ with respect to $a$. $(a,b)\sim U\left\{(x,y):x^2+y^2\leq r^2\right\}$ represents that the point $(a,b)$ is uniformly distributed in the circular region with center of $(0,0)$ and radius of $r$.

\section{System Model and Working Process Design}

\begin{figure}[!t]
\includegraphics[width=3.6in]{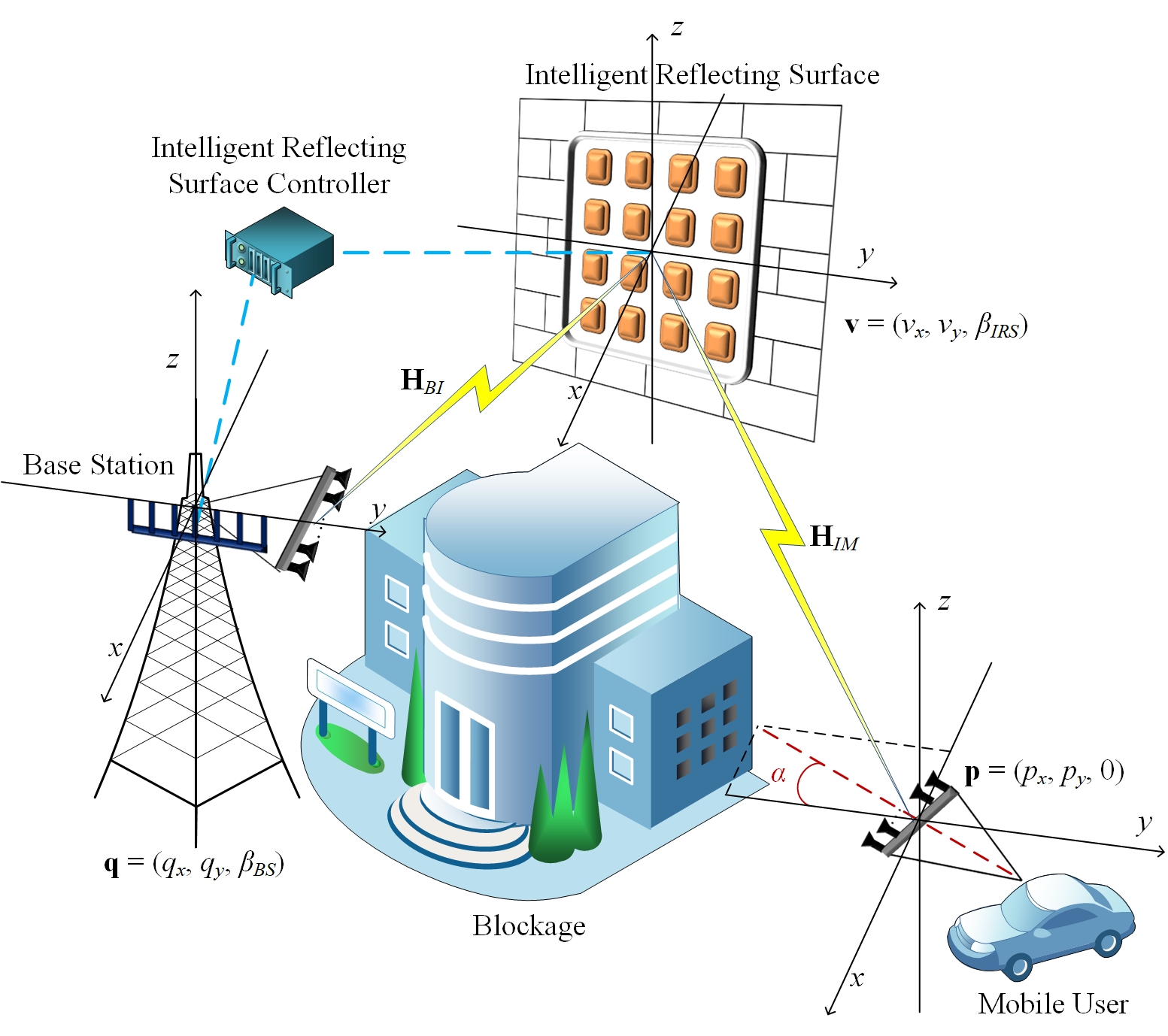}
\hfil
\centering
\caption{The considered IMM-JLCS. A multiple-antenna BS localizes and communicates with a multiple-antenna MU with the aid of an IRS, when the LoS path is blocked by an obstacle.}
\label{IRS localization--blockage-system}
\end{figure}

We consider an IMM-JLCS in the 3D scenario, as illustrated in Figure \ref{IRS localization--blockage-system}. A BS with uniform linear array (ULA) consisting of $N_B^t$ antennas, localizes and communicates with a MU with ULA consisting of $N_M^r$ antennas. The LoS path is assumed to be obstructed by an obstacle, e.g. the edifice or infrastructure. Due to the mmWave's susceptibility to blockage, the LoS link is assumed unavailable in such situations, so that a uniform square planar IRS containing $N=L\times L$ passive reflecting elements, with $L$ being the number of rows or columns of the IRS, is employed to establish a strong VLoS reflection path to assist the localization and wireless data transmission. The adjacent element spacing of the antennas on the BS/MU or the reflecting units on the IRS is $d=\lambda/2$, where $\lambda$ denotes the signal wavelength. 
To facilitate the analysis, an auxiliary 3D Cartesian coordinate system is established to indicate the positions of the IRS, the BS and the MU. The IRS and the antenna array on the BS are placed parallel to $y-o-z$ plane and $x$-axis, respectively, with their centers located at $\mathbf{v}=(v_x,v_y,\beta_{IRS})^T\in \mathbb{R}^3$ and $\mathbf{q}=(q_x,q_y,\beta_{BS})^T\in \mathbb{R}^3$, where $\beta_{IRS}$ and $\beta_{BS}$ symbolize the heights of the IRS and BS relative to the MU on the ground. The MU moves and rotates by angle $\alpha\in[0,2\pi)$ in the $x-o-y$ plane, with the center of its antenna array located at $\mathbf{p}=(p_x,p_y,0)^T\in \mathbb{R}^3$. Here, $\mathbf{v}$ and $\mathbf{q}$ are known and invariant after the deployments of the IRS and BS, while $\mathbf{p}$ and $\alpha$ are unknown and need to be estimated. 

\begin{figure}[!t]
\includegraphics[width=6.6in]{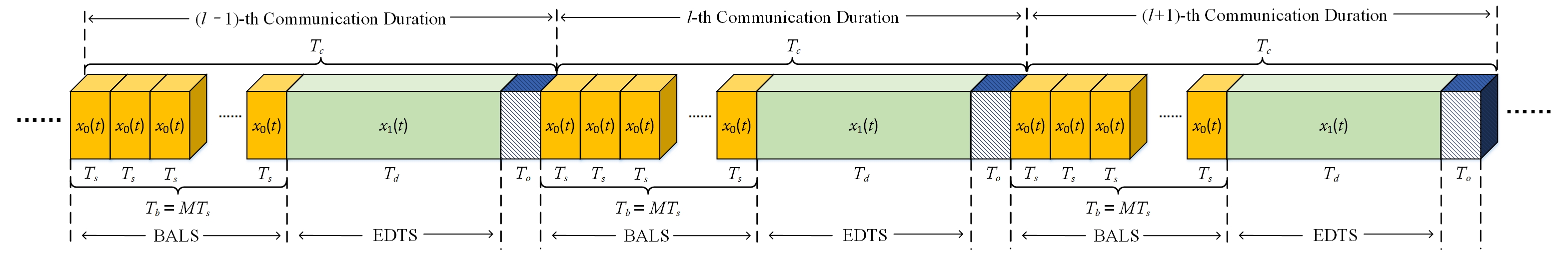}
\hfil
\centering
\caption{Timeline of the localization, communication and joint optimization. Each communication period has a fixed duration of $T_c$, and involves three stages, i.e. the BALS, EDTS and JOS, with the durations of $T_b$, $T_d$ and $T_o$, respectively. $M$ stands for the number of the transmitted pilot signals with the duration of $T_s$ in one communication period, and satisfies $M=N_B\times N_M$.}
\label{Time}
\end{figure}

This system is designed to achieve the goals of: 1) localizing the MU and determining its orientation from the received signals; 2) transmitting effective data from the BS to the MU; and 3) jointly optimizing the position/orientation estimation accuracy and the EADR. The three tasks are completed independently in a communication period with the fixed duration of $T_c$. Specifically, as shown in Figure \ref{Time}, which indicates the timeline of the tasks, one communication period is divided into three stages: the BALS with the duration of $T_{b}$, the EDTS with the duration of $T_{d}$, and the JOS with the duration of $T_{o}$. In the BALS, the BS sequentially emits several pilot signals to the MU for beam alignment and position/orientation estimation. Afterwards, in the EDTS, the BS communicates with the MU by transmitting the information-carrying signal. Finally, in the JOS, the system performs joint optimization on both the localization and communication performances. When the system is running, $T_{c}$ and $T_{o}$ are fixed, while $T_{b}$ and $T_{d}$ are alterable but satisfy $T_{b}+T_{d}=T_{c}-T_{o}$. The time allocation ratio for $T_b$ is denoted by $\varpi=\frac{T_b}{T_c}$, while that for $T_d$ is $\left(1-\frac{T_o}{T_c}-\varpi\right)$. Moreover, the BS can partially activate $N_B\leq N_B^t$ adjacent antennas for signal emission and deactivate the other $N_B^t-N_B$ antennas, while the MU activates totally $N_M=N_M^r$ antennas for signal reception. The position of the MU is assumed to be approximately invariant within one communication period, but change between distinct communication periods. The MU's position in the $(l-1)$-th communication period, denoted by $(p_x,p_y)|_{(l-1)}$, is uniformly distributed within a circular area with the radius of $\Upsilon_{(x,y)}$ and center point of the MU's position in the $l$-th communication period, denoted by $(p_x,p_y)|_{l}$, i.e. $(p_x,p_y)|_{(l-1)}=(p_x,p_y)|_{l}+(\delta_{p_x},\delta_{p_y})$, with $(\delta_{p_x},\delta_{p_y})$ given by

\begin{equation}
(\delta_{p_x},\delta_{p_y})\sim U\left\{(x,y):x^2+y^2\leq \Upsilon_{(x,y)}^2\right\}
\end{equation}
where different $\Upsilon_{(x,y)}$ can embody different levels of user mobility.

Based on the aforementioned geometry and functionality of this system, we will subsequently illustrate the signal, channel and estimation error models, and elaborate the system's working process in detail.

\subsection{Transmitted Signal Model}
We first describe the transmitted signal models in the BALS and EDTS. In the BALS, let the pilot signal be denoted by a continuous time-domain waveform $x_0(t)$, with the bandwidth of $B$ and duration of $T_s$. For simple analysis, we assume that $x_0(t)$ has the unit power and flat spectrum, which causes its power spectrum, denoted by $|X_0(\omega)|^2$, to be a constant of $|X_0(\omega)|^2=\frac{T_s}{2\pi B}$ within $\omega\in[-\pi B,\pi B]$, where $X_0(\omega)=\int_{-\infty}^{\infty} x_0(t)e^{-j\omega t}dt$ is the Fourier transform of $x_0(t)$ \cite{Destino2017(ICC)}. When transmitting each pilot signal, the BS selects a column of codewords as the unit-norm transmit beamforming vector $\mathbf{w}_B$, which satisfies $\|\mathbf{w}_B\|=1$, from a predetermined DFT codebook $\bm{\mathcal{C}}_{BS}\in \mathbb{C}^{N_B\times N_B}$. Specifically, the $m_B$-th column of $\bm{\mathcal{C}}_{BS}$ is expressed as
\begin{equation}\label{m_B_column}
\left[\bm{\mathcal{C}}_{BS}\right]_{m_B}=\frac{1}{\sqrt{N_B}}\left(1,e^{-j\frac{2\pi}{N_B}(m_B-1)},...,e^{-j\frac{2\pi}{N_B}(m_B-1)(N_B-1)}\right)^T
\end{equation}
where $m_B=1,2,...,N_B$.
Thus, the transmitted pilot signal is expressed as
\begin{equation}\label{transmit pilot signal}
\mathbf{x}_0(t)=\sqrt{P_{TX}}\mathbf{w}_B x_0(t)
\end{equation}
where $\sqrt{P_{TX}}$ stands for the average transmitting power.

In the EDTS, let the signal carrying the effective information be denoted by $x_1(t)$ with the bandwidth of $B$ and the duration of $T_d$. Hence, the transmitted information-carrying signal is expressed as
\begin{equation}\label{transmit effective signal}
\mathbf{x}_1(t)=\sqrt{P_{TX}}\widetilde{\mathbf{w}}_B x_1(t)
\end{equation}
where $\widetilde{\mathbf{w}}_B$ represents the optimal transmit beamformer, which, together with the optimal receive combining vector $\widetilde{\mathbf{w}}_M$ at the MU (detailed in Section II-C), contributes to maximizing the received signal-to-noise ratio (SNR) among all the beamformers searched from $\bm{\mathcal{C}}_{BS}$. 
%This will be detailed in Section II-D: \textit{Working Process}.

\subsection{Wireless Channel Model}
We next illustrate the wireless channel model. As shown in Figure 1, the VLoS channel is composed of two tandem parts, denoted by $\mathbf{H}_{BI}$ from the BS to the IRS, and $\mathbf{H}_{IM}$ from the IRS to the MU, which are modelled as
\begin{equation}\label{channel BS-IRS}
\mathbf{H}_{BI}=\mathbf{a}_{IRS}(\varphi_{IRS,1}^a,\varphi_{IRS,1}^e)\mathbf{a}_{TX}^H(\varphi_{TX,1})
\end{equation}
\begin{equation}\label{channel IRS-MU}
\mathbf{H}_{IM}=\mathbf{a}_{RX}(\varphi_{RX,1})\mathbf{a}_{IRS}^H(\varphi_{IRS,2}^a,\varphi_{IRS,2}^e)
\end{equation}
where $\varphi_{IRS,1}^a$, $\varphi_{IRS,1}^e$ and $\varphi_{TX,1}$ are the azimuth AOA, elevation AOA at the IRS and the AOD at the BS for the BS-IRS link, while $\varphi_{IRS,2}^a$, $\varphi_{IRS,2}^e$ and $\varphi_{RX,1}$ are the azimuth AOD, elevation AOD at the IRS and the AOA at the MU for the IRS-MU link. These parameters are related to the positions and rotation angle of the MU according to
\begin{equation}\label{Relation4}
\varphi_{TX,1}=arcsin\left(\frac{v_x-q_x}{\|\mathbf{v}-\mathbf{q}\|_2}\right)
\end{equation}
\begin{equation}\label{Relation5}
\varphi_{IRS,1}^a=arcsin\left(\frac{v_y-q_y}{\sqrt{(v_x-q_x)^2+(v_y-q_y)^2}}\right)
\end{equation}
\begin{equation}\label{Relation6}
\varphi_{IRS,2}^a=arcsin\left(\frac{p_y-v_y}{\sqrt{(p_x-v_x)^2+(p_y-v_y)^2}}\right)
\end{equation}
\begin{equation}\label{Relation7}
\varphi_{IRS,1}^e=arccos\left(\frac{\beta_{IRS}-\beta_{BS}}{\|\mathbf{v}-\mathbf{q}\|_2}\right)
\end{equation}
\begin{equation}\label{Relation8}
\varphi_{IRS,2}^e=arccos\left(\frac{\beta_{IRS}}{\|\mathbf{p}-\mathbf{v}\|_2}\right)
\end{equation}
\begin{equation}\label{Relation10}
\varphi_{RX,1}=arcsin\left(\frac{(p_x-v_x)cos\alpha-(p_y-v_y)sin\alpha}{\|\mathbf{p}-\mathbf{v}\|_2}\right)
\end{equation}

The array response vectors in $\mathbf{H}_{BI}$ and $\mathbf{H}_{IM}$ are given by
\begin{equation}\label{aTX1}
\mathbf{a}_{TX}(\varphi_{TX,1})=\left(1,e^{j\frac{2\pi d}{\lambda}sin\varphi_{TX,1}},...,e^{j\frac{2\pi d}{\lambda}(N_B-1)sin\varphi_{TX,1}}\right)^T
\end{equation}
\begin{equation}\label{aRX1}
\mathbf{a}_{RX}(\varphi_{RX,1})=\left(1,e^{j\frac{2\pi d}{\lambda}sin\varphi_{RX,1}},...,e^{j\frac{2\pi d}{\lambda}(N_M-1)sin\varphi_{RX,1}}\right)^T
\end{equation}
\begin{equation}\label{aIRS1}
\begin{split}
\mathbf{a}_{IRS}(\varphi_{IRS,1}^a,\varphi_{IRS,1}^e)=&\left(1,e^{j\frac{2\pi d}{\lambda} cos\varphi_{IRS,1}^e},...,e^{j\frac{2\pi d}{\lambda}(L-1) cos\varphi_{IRS,1}^e}\right)^T \otimes\\
&\left(1,e^{j\frac{2\pi d}{\lambda}sin\varphi_{IRS,1}^a sin\varphi_{IRS,1}^e},...,e^{j\frac{2\pi d}{\lambda}(L-1)sin\varphi_{IRS,1}^a sin\varphi_{IRS,1}^e}\right)^T
\end{split}
\end{equation}
\begin{equation}\label{aIRS2}
\begin{split}
\mathbf{a}_{IRS}(\varphi_{IRS,2}^a,\varphi_{IRS,2}^e)=
&\left(1,e^{j\frac{2\pi d}{\lambda} cos\varphi_{IRS,2}^e},...,e^{j\frac{2\pi d}{\lambda}(L-1) cos\varphi_{IRS,2}^e}\right)^T \otimes\\
&\left(1,e^{j\frac{2\pi d}{\lambda}sin\varphi_{IRS,2}^a sin\varphi_{IRS,2}^e},...,e^{j\frac{2\pi d}{\lambda}(L-1)sin\varphi_{IRS,2}^a sin\varphi_{IRS,2}^e}\right)^T
\end{split}
\end{equation}

As the IRS is able to induce adjustable phase shifts on the impinging signal wave from the BS through $\mathbf{H}_{BI}$, and reflect it to the MU through $\mathbf{H}_{IM}$, the entire channel is expressed as
\begin{equation}\label{reflect channel}
\mathbf{H}_{BIM}=\widetilde{h}_1\mathbf{H}_{IM} \mathbf{\Theta} \mathbf{H}_{BI}
\end{equation}
where $\widetilde{h}_1=\frac{h_1}{\sqrt{\rho_1}}$, with $h_1$ symbolizing the complex channel gain, and $\rho_1$ embodying the path loss of the BS-IRS-MU link. Specifically, $\rho_1$ satisfies $\frac{1}{\rho_{1}}=\zeta^2\left[\frac{\lambda}{4\pi (d_{1,1}+d_{1,2})}\right]^2$ \cite{A.Elzanaty2020(ArXiv)}, where $\zeta$ denotes the power attenuation coefficient; $d_{1,1}=\|\mathbf{v}-\mathbf{q}\|_2$ and $d_{1,2}=\|\mathbf{p}-\mathbf{v}\|_2$ are the distances between the BS and IRS, and between the IRS and MU, respectively. $\mathbf{\Theta}=\delta\times diag(e^{j\theta_1},e^{j\theta_2},...,e^{j\theta_N})\in \mathbb{C}^{N\times N}$ is the diagonal phase shifting matrix of the IRS, in which $\delta\in(0,1]$ represents the reflection coefficient, and $\theta_i$, for $i=1,2,...,N$, represents the $i$-th IRS phase shift. In the considered system, $\mathbf{\Theta}$ can be configured differently in the BALS and EDTS. Specifically, in the BALS, $\mathbf{\Theta}$ can be adjusted into $\mathbf{\Theta}=\widetilde{\mathbf{\Theta}}_1$, which is the optimal configuration for the localization performance. In the EDTS, $\mathbf{\Theta}$ can be adjusted into $\mathbf{\Theta}=\widetilde{\mathbf{\Theta}}_2$, which is the optimal configuration for the data transmission performance. Detailed information about the configurations of $\widetilde{\mathbf{\Theta}}_1$ and $\widetilde{\mathbf{\Theta}}_2$ will be given in Section III.

\subsection{Received Signal Model}
We then elaborate the received signal models in the BALS and EDTS. In the BALS, when the MU receipts one pilot signal, it selects a column of codewords as the receive combining vector $\mathbf{w}_M\in\mathbb{C}^{N_M}$, which satisfies $\|\mathbf{w}_M\|=1$, from a predetermined DFT codebook $\bm{\mathcal{C}}_{MU}\in \mathbb{C}^{N_M\times N_M}$. 
Specifically, the $m_M$-th column of $\bm{\mathcal{C}}_{MU}$ is expressed as
\begin{equation}\label{m_M_column}
\left[\bm{\mathcal{C}}_{MU}\right]_{m_M}=\frac{1}{\sqrt{N_M}}\left(1,e^{-j\frac{2\pi}{N_M}(m_M-1)},...,e^{-j\frac{2\pi}{N_M}(m_M-1)(N_M-1)}\right)^T
\end{equation}
where $m_M=1,2,...,N_M$.
Hence, the received pilot signal is expressed as
\begin{equation}\label{received pilot signal}
\begin{split}
y_0(t)=\widetilde{h}_1\mathbf{w}_M^H\mathbf{H}_{IM}\widetilde{\mathbf{\Theta}}_1\mathbf{H}_{BI}\mathbf{x}_0(t\!-\!\tau_1)+\mathbf{w}_M^H\mathbf{n}(t)
\end{split}
\end{equation}
where $\tau_1=\frac{\|\mathbf{v}-\mathbf{q}\|_2+\|\mathbf{p}-\mathbf{v}\|_2}{c}$ symbolizes the time delay of the VLoS path; $c\approx 2.99792458\times 10^8\ m/s$ is the speed of light. $\mathbf{n}(t)$ stands for the additive white Gaussian noise (AWGN) at the MU, with mean of 0, variance of $\sigma_w^2$ and power spectral density of $N_0=\frac{\sigma_w^2}{B}$.

In the EDTS, the received information-carrying signal is expressed as
\begin{equation}\label{received effective signal}
\begin{split}
y_1(t)=\widetilde{h}_1\widetilde{\mathbf{w}}_M^H\mathbf{H}_{IM}\widetilde{\mathbf{\Theta}}_2\mathbf{H}_{BI}\mathbf{x}_1(t\!-\!\tau_1)+\widetilde{\mathbf{w}}_M^H\mathbf{n}(t)
\end{split}
\end{equation}
where $\widetilde{\mathbf{w}}_M$ is the optimal receive combining vector, which, together with the optimal transmit beamformer $\widetilde{\mathbf{w}}_B$ at the BS, contributes to maximizing the received SNR among all the receive combining vectors searched from $\bm{\mathcal{C}}_{MU}$. 
%This will be detailed in Section II-D: \textit{Working Process}.

\subsection{Working Process}

\begin{figure}[!t]
\includegraphics[width=6.6in]{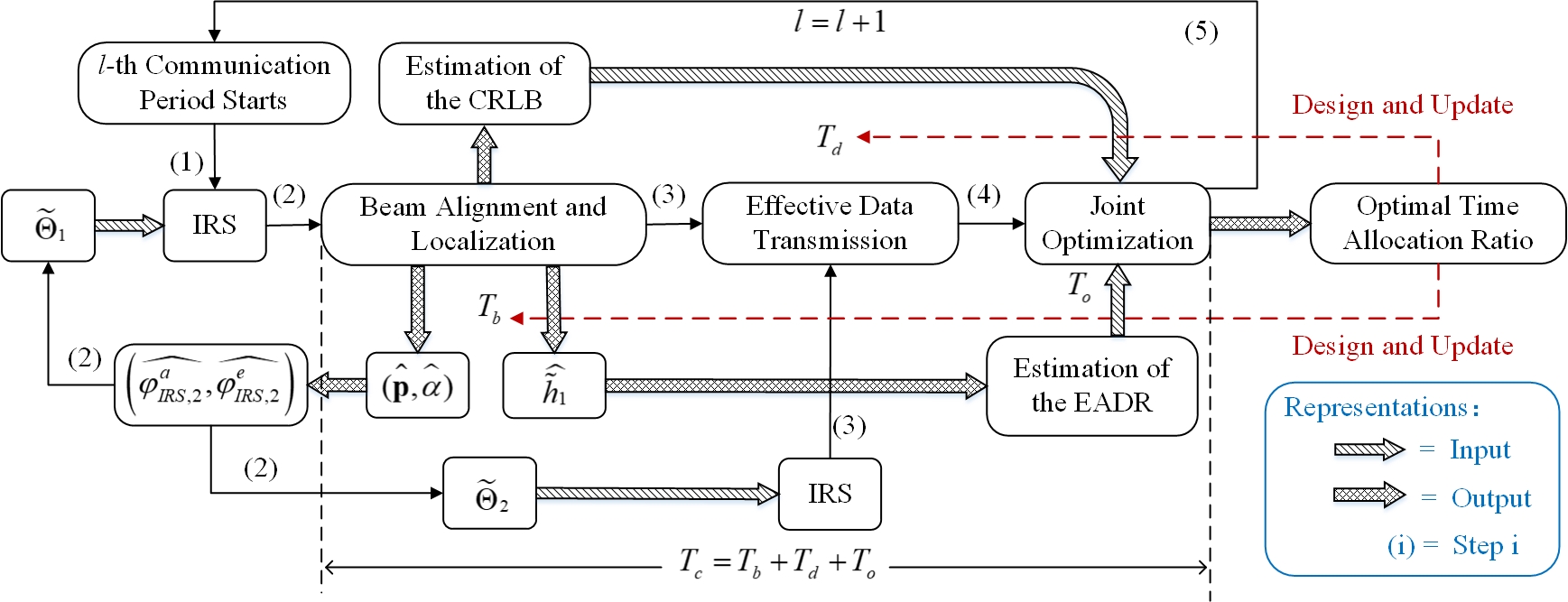}
\hfil
\centering
\caption{Schematic of the working process of the IRS-aided joint localization and communication system in the $l$-th communication period.}
\label{Working Process}
\end{figure}

We subsequently expound the working process of this system. Here, we consider the $l$-th communication period as an example, and present the flowchart of the working process in Figure \ref{Working Process}. From Figure \ref{Working Process}, we describe the procedure as five steps:

\begin{itemize}
\item \textit{Step 1}: When the $l$-th communication period begins, the IRS controller determines a $\widetilde{\mathbf{\Theta}}_1$ based on the estimated azimuth AOD ($\widehat{\varphi_{IRS,2}^a}$) and elevation AOD ($\widehat{\varphi_{IRS,2}^e}$) at the IRS from the $(l-1)$-th communication period, and adjusts the IRS phase shifting matrix into $\widetilde{\mathbf{\Theta}}_1$.

\item \textit{Step 2}: The BS and MU simultaneously search $\mathbf{w}_B$ and $\mathbf{w}_M$ from $\bm{\mathcal{C}}_{BS}$ and $\bm{\mathcal{C}}_{MU}$ column-by-column, i.e. \textit{exhaustive search}, in the BALS. When searching each beam pair of $(\mathbf{w}_B,\mathbf{w}_M)$, the BS transmits a pilot signal $x_0(t)$ to the MU for beam alignment and position/orientation estimation. When this procedure ends, the BS totally sends $M=N_B\times N_M$ pilot signals, after which it outputs the estimated parameters including $\widehat{\widetilde{h}_1}$, $\widehat{\mathbf{p}}$, $\widehat{\alpha}$, as well as $\widehat{\varphi_{IRS,2}^a}$ and $\widehat{\varphi_{IRS,2}^e}$ which are mapped from $\widehat{\mathbf{p}}$ and $\widehat{\alpha}$ according to (\ref{Relation6}) and (\ref{Relation8}), and then calculates the CRLB of the position/orientation estimation error based on (\ref{closed-form PREB}) in Section IV-B. The $\widehat{\varphi_{IRS,2}^a}$ and $\widehat{\varphi_{IRS,2}^e}$ are stored for determining $\widetilde{\mathbf{\Theta}}_1$ in the next communication period, and are adopted by the IRS controller to determine $\widetilde{\mathbf{\Theta}}_2$ for the subsequent effective data transmission in this communication period. The $\widehat{\widetilde{h}_1}$ is substituted into (\ref{closed-form EADR}) in Section IV-B for calculating the EADR of this communication period. The CRLB and EADR are stored as objectives to be optimized, and will be input to the joint optimization module in \textit{Step 4}.  

\item \textit{Step 3}: When the BALS terminates, the BS and MU decide a beam pair of $(\widetilde{\mathbf{w}}_M,\widetilde{\mathbf{w}}_B)$, which is selected corresponding to the maximum received SNR from all beam pairs searched during the BALS.
Then, the IRS phase shifting matrix is adjusted into $\widetilde{\mathbf{\Theta}}_2$ in the EDTS, and the BS sends $x_1(t)$ to the MU for effective data transmission.

\item \textit{Step 4}: When the EDTS terminates, the signal transmission and reception are suspended, and the joint optimization module is actuated. The CRLB and EADR, obtained in \textit{Step 2}, are input to the joint optimization module, which aims at finding a $\varpi$ that makes the CRLB and EADR jointly optimal. After the optimization process, the output of $\varpi$ is used to design and update $T_b$ and $T_d$, which guides the BS to determine the number of the activated antennas or the codebook size, for the next communication period. 

%\begin{equation}\label{EADR}
%\begin{split}
%R_{eff}=\left(1-\frac{MT_s+T_o}{T_c}\right)
%B\times\log_2\left(1+\frac{P_{TX}|\widehat{\widetilde{h}_1}|^2|\widetilde{\mathbf{w}}_M^H\widehat{\mathbf{H}_{IM}} \widetilde{\mathbf{\Theta}}_2 \mathbf{H}_{BI}\widetilde{\mathbf{w}}_B|^2}{N_0B}\right)
%\end{split}
%\end{equation}
%where $\widehat{\mathbf{H}_{IM}}$ is expressed as
%\begin{equation}\label{channel IRS-MU estimated}
%\widehat{\mathbf{H}_{IM}}=\mathbf{a}_{RX}(\widehat{\varphi_{RX,1}})\mathbf{a}_{IRS}^H(\widehat{\varphi_{IRS,2}^a},\widehat{\varphi_{IRS,2}^e})
%\end{equation}
%in which $\widehat{\varphi_{RX,1}}$ is mapped from $\widehat{\mathbf{p}}$ and $\widehat{\alpha}$ according to (\ref{Relation10}).

\item \textit{Step 5}: The working process turns to the beginning of the $(l+1)$-th communication period, and repeats \textit{Step 1} to \textit{Step 4}.

\end{itemize}

It is noted that the $\widetilde{\mathbf{\Theta}}_1$ in the $l$-th communication period is determined by $\widehat{\varphi_{IRS,2}^a}$ and $\widehat{\varphi_{IRS,2}^e}$, which are estimated in the $(l-1)$-th communication period. This may result in a hysteretic update of $\widetilde{\mathbf{\Theta}}_1$ in the presence of user mobility, which will be discussed in the simulations.

In addition, as the positioning and channel estimation algorithms in the mmWave communication scenarios have been investigated in the related works \cite{mmWave MIMO Localization, Shahmansoori2018(TWC), Wang2019(TSP), H.Zhang2020(ArXiv)}, here we assume that $\widehat{\widetilde{h}_1}$, $\widehat{\mathbf{p}}$ and $\widehat{\alpha}$ can be acquired by some existing estimation techniques based on, e.g. maximum-likelihood, compressed sensing, \textit{et al.} which are out of the scope of this paper. Instead, we model the estimation errors below, which exist in most estimation methods.

\subsection{Estimation Error Model}

\subsubsection{Localization Error}

By referring to \cite{X.Hu2020(TCOM)}, the estimated position $\widehat{\mathbf{p}}$ and rotation angle $\widehat{\alpha}$ are, respectively, expressed as
\begin{equation}
\widehat{\mathbf{p}}=\mathbf{p} + \Delta\mathbf{p}
\end{equation}
\begin{equation}
\widehat{\alpha}=\alpha + \Delta\alpha
\end{equation}
where $\mathbf{p}$ and $\alpha$ are the actual position and rotation angle of the MU; $\Delta\mathbf{p}$ represents the position estimation error, which is uniformly distributed within a circular region with the radius of $\epsilon_{(x,y)}$ and center point of $(0,0)$; $\Delta\alpha$ represents the rotation estimation error, which is uniformly distributed within $[-\epsilon_\alpha,\epsilon_\alpha]$.

\subsubsection{Channel Estimation Error}

According to \cite{J.Zhang2020(CL)}, the estimated complex channel coefficient $\widehat{\widetilde{h}_1}$ is expressed as
\begin{equation}
\widehat{\widetilde{h}_1}=\widetilde{h}_1+\Delta \widetilde{h}_1
\end{equation}
where $\widetilde{h}_1$ is the actual complex channel coefficient; $\Delta \widetilde{h}_1$ denotes the channel estimation error, which follows a zero-mean complex Gaussian distribution with the variance of $\sigma_h^2$.

Based on the signal and estimation error models, in the next section, we will first obtain the position error bound (PEB), rotation error bound (REB) and EADR, and then derive their closed-form expressions with respect to the time allocation ratio $\varpi$.

\section{System Performance Metrics}
For evaluating the performances of the position/orientation estimation and effective data transmission, this section first introduces PEB/REB and EADR as performance metrics, and then derives their closed-form expressions in relation to $\varpi$ by configuring appropriate IRS phase shifting matrices in the BALS and EDTS.
%in order to facilitate the following joint optimization approach in Section IV.

\subsection{PEB and REB}

\subsubsection{Mathematical Description}

$\ $

The PEB and REB can be derived from the Fisher information matrix (FIM). Preceding the derivations of the FIM, we should first define a variable vector containing the unknown channel parameters to be estimated. As described in Section II, because the positions of the BS and IRS are known, $\varphi_{TX,1}$, $\varphi_{IRS,1}^a$ and $\varphi_{IRS,1}^e$ can be geometrically determined by $\mathbf{q}$ and $\mathbf{v}$. As a result, the unknown parameters are those related to $\mathbf{p}$, which are involved in
\begin{equation}\label{nita}
\begin{split}
\bm{\eta}=(\tau_1,\varphi_{RX,1},\varphi_{IRS,2}^a,
\varphi_{IRS,2}^e,\widetilde{h}_{\mathfrak{R},1},\widetilde{h}_{\mathfrak{I},1})^T \in\mathbb{R}^{6}
\end{split}
\end{equation}
where $\widetilde{h}_{\mathfrak{R},1}=\mathfrak{Re}\{\widetilde{h}_1\}$ and $\widetilde{h}_{\mathfrak{I},1}=\mathfrak{Im}\{\widetilde{h}_1\}$ are the real part and imaginary part of $\widetilde{h}_1$.

The FIM of $\bm{\eta}$ with respect to $\mathbf{w}_B$, $\mathbf{w}_M$ and $\widetilde{\mathbf{\Theta}}_1$ can be derived from \cite{mmWave MIMO Localization}:
\begin{equation}\label{J_eta}
\mathbf{J}_{\bm{\eta}}(\mathbf{w}_B,\widetilde{\mathbf{\Theta}}_1,\mathbf{w}_M)=
\mathbb{E}_{y_0|\bm{\eta}}\left[-\frac{\partial^{2}\ln f(y_0|\bm{\eta})}{\partial\bm{\eta}\partial\bm{\eta}^T}\right]
\end{equation}
where $f(y_0|\bm{\eta})$ is the likelihood function of $y_0(t)$ conditioned on $\bm{\eta}$, given by
\begin{equation}\label{f}
f(y_0|\bm{\eta})\propto exp\left\{\frac{2}{N_0}\int_0^{T_s}u_0^*(t)y_0(t)dt-\frac{1}{N_0}\int_0^{T_s}|u_0(t)|^2dt\right\}
\end{equation}
where
\begin{equation}\label{u_0(t)}
u_0(t)=\widetilde{h}_1\mathbf{w}_M^H\mathbf{H}_{IM}\widetilde{\mathbf{\Theta}}_1\mathbf{H}_{BI}\mathbf{x}_0(t\!-\!\tau_1)
\end{equation}
is the deterministic noiseless signal part in $y_0(t)$.

From (\ref{J_eta}) and (\ref{f}), after a few manipulations, $\mathbf{J}_{\bm{\eta}}(\mathbf{w}_B,\widetilde{\mathbf{\Theta}}_1,\mathbf{w}_M)$ is simplified into
\begin{equation}\label{J_eta2}
\mathbf{J}_{\bm{\eta}}(\mathbf{w}_B,\widetilde{\mathbf{\Theta}}_1,\mathbf{w}_M)=
\frac{1}{N_0}\int_0^{T_s}\mathfrak{Re}\left\{\bigtriangledown_{\bm{\eta}}^Hu_0(t)\bigtriangledown_{\bm{\eta}}u_0(t)\right\}dt
\end{equation}
whose $(i,j)$-th entry is given by
\begin{equation}\label{J_eta3}
J_{\eta_i,\eta_j}=
\frac{1}{N_0}\int_0^{T_s}\mathfrak{Re}\left\{\frac{\partial u_0^*(t)}{\partial\eta_i}\times \frac{\partial u_0(t)}{\partial\eta_j}\right\}dt
\end{equation}
where $\eta_k$, for $k=1,2,...,6$, is the $k$-th parameter in $\bm{\eta}$.

In order to avoid affecting the coherence of reading, we provide the derivations and exact expressions of the elements in $\mathbf{J}_{\bm{\eta}}(\mathbf{w}_B,\widetilde{\mathbf{\Theta}}_1,\mathbf{w}_M)$ in Appendix A. 

Afterwards, we calculate the FIM for $p_x$, $p_y$ and $\alpha$.
Let $\mathbf{T}\in\mathbb{R}^{3\times6}$ denote a Jacobian matrix composed of the partial derivatives of the channel parameters in $\bm{\eta}$ on $p_x$, $p_y$ and $\alpha$. The elements in $\mathbf{T}$ are derived in Appendix B.

Finally, let $\mathbf{g}$ be defined by $\mathbf{g}=(p_x,p_y,\alpha)^T$. Thus, the FIM for $\mathbf{g}$, defined by $\mathbf{J}_\mathbf{g}(\mathbf{w}_B,\widetilde{\mathbf{\Theta}}_1,\mathbf{w}_M)$, is expressed as
\begin{equation}\label{J_g}
\mathbf{J}_\mathbf{g}(\mathbf{w}_B,\widetilde{\mathbf{\Theta}}_1,\mathbf{w}_M)=\mathbf{T}\mathbf{J}_{\bm{\eta}}(\mathbf{w}_B,\widetilde{\mathbf{\Theta}}_1,\mathbf{w}_M)\mathbf{T}^T
\end{equation}

It is notable that $\mathbf{J}_\mathbf{g}(\mathbf{w}_B,\widetilde{\mathbf{\Theta}}_1,\mathbf{w}_M)$ in (\ref{J_g}) is computed for a single beam pair of $(\mathbf{w}_B,\mathbf{w}_M)$. As there are $M=N_B\times N_M$ beam pairs searched during the beam alignment procedure and the FIM is calculated for all the beam pairs, we will totally obtain $M$ different $\mathbf{J}_\mathbf{g}(\mathbf{w}_B,\widetilde{\mathbf{\Theta}}_1,\mathbf{w}_M)$ in the BALS. Owing to the additive property of the FIM, we obtain the entire FIM according to
\begin{equation}\label{Sum_J}
\mathbf{J}_{\sum_{}^{}}=\sum_{(\mathbf{w}_B,\widetilde{\mathbf{\Theta}}_1,\mathbf{w}_M)}\mathbf{J}_\mathbf{g}(\mathbf{w}_B,\widetilde{\mathbf{\Theta}}_1,\mathbf{w}_M)
\end{equation}

From (\ref{Sum_J}), we obtain the actual PEB in meters and REB in radians for the MU by calculating
\begin{equation}\label{PEB}
PEB=\sqrt{tr\left\{\left[\mathbf{J}_{\sum_{}^{}}^{-1}\right]_{1:2,1:2}\right\}}=\min{\left\{\sqrt{var(\widehat{\mathbf{p}})}\right\}}
\end{equation}
\begin{equation}\label{REB}
REB=\sqrt{tr\left\{\left[\mathbf{J}_{\sum_{}^{}}^{-1}\right]_{3,3}\right\}}=\min{\left\{\sqrt{var(\widehat{\alpha})}\right\}}
\end{equation}
where $\left[\mathbf{J}_{\sum_{}^{}}^{-1}\right]_{a:b,c:d}$ represents the submatrix constructed by the $a$-th to $b$-th rows and the $c$-th to $d$-th columns of $\mathbf{J}_{\sum_{}^{}}^{-1}$; $var(.)$ denotes the variance of the random variable inside $(.)$.

Eq. (\ref{PEB}) and Eq. (\ref{REB}) indicate that the PEB and REB are proportional to the trace of the inverse matrix of $\mathbf{J}_{\sum_{}^{}}$, so that the more beams are searched during the beam alignment, the more $\mathbf{J}_\mathbf{g}(\mathbf{w}_B,\widetilde{\mathbf{\Theta}}_1,\mathbf{w}_M)$ and the lower PEB or REB are obtained, which improves the potential location/orientation estimation performance. 

$\ $

\subsubsection{Phase Shift Design}

$\ $

Based on (\ref{PEB}) and (\ref{REB}), the optimal configuration of $\widetilde{\mathbf{\Theta}}_1$ in the $l$-th communication period should make the FIM achieve its maximum. Referring to \textbf{Observation 2} and Eq. (12) in \cite{J.He2020(VTC)}, we similarly retrospect $\gamma_{IRS}$ in Appendix A, on which the position/rotation estimation performance primarily depends if $\mathbf{w}_B$ and $\mathbf{w}_M$ are given. Because $\gamma_{IRS}$ can be further expressed as
\begin{equation}\label{gamma_IRS_transform}
\begin{split}
\gamma_{IRS}=&\mathbf{a}_{IRS}^H(\varphi_{IRS,2}^a,\varphi_{IRS,2}^e)\widetilde{\mathbf{\Theta}}_1\mathbf{a}_{IRS}(\varphi_{IRS,1}^a,\varphi_{IRS,1}^e)\\
=&\left[\mathbf{a}_{IRS}(\varphi_{IRS,2}^a,\varphi_{IRS,2}^e)\odot\mathbf{a}_{IRS}^*(\varphi_{IRS,1}^a,\varphi_{IRS,1}^e)\right]^H \widetilde{\bm{\theta}}_1
\end{split}
\end{equation}
where $\widetilde{\bm{\theta}}_1$ is a column vector which satisfies $\widetilde{\mathbf{\Theta}}_1=diag\left(\widetilde{\bm{\theta}}_1\right)$. According to (\ref{gamma_IRS_transform}), we have $|\gamma_{IRS}|\leq N$, and $|\gamma_{IRS}|$ reaches its maximum when $\widetilde{\bm{\theta}}_1=\mathbf{a}_{IRS}(\varphi_{IRS,2}^a,\varphi_{IRS,2}^e)\odot\mathbf{a}_{IRS}^*(\varphi_{IRS,1}^a,\varphi_{IRS,1}^e)$. Because the IRS phase shifts are adjusted based on the estimated parameters, the optimal $\widetilde{\mathbf{\Theta}}_1$ is designed as 
\begin{equation}
\widetilde{\mathbf{\Theta}}_1=diag\left(\mathbf{a}_{IRS}(\widehat{\varphi_{IRS,2}^a}(l-1),\widehat{\varphi_{IRS,2}^e}(l-1))\odot\mathbf{a}_{IRS}^*(\varphi_{IRS,1}^a,\varphi_{IRS,1}^e)\right)
\end{equation}
where $\widehat{\varphi_{IRS,2}^a}(l-1)$ and $\widehat{\varphi_{IRS,2}^e}(l-1)$, according to Figure \ref{Working Process}, represent the estimated $\varphi_{IRS,2}^a$ and $\varphi_{IRS,2}^e$ in the $(l-1)$-th communication period.

$\ $

\subsubsection{Approximate Closed-form Expression}

$\ $

From Appendix A, the elements in $\mathbf{J}_{\bm{\eta}}(\mathbf{w}_B,\widetilde{\mathbf{\Theta}}_1,\mathbf{w}_M)$ are related to $\gamma_{TX,1}$, $\gamma_{RX,1}$,$\gamma_{IRS}$, $\overline{\gamma_{RX,1}}$, $\overline{\gamma_{IRS,a}}$ and $\overline{\gamma_{IRS,e}}$, if $P_{TX}$, $T_s$, $N_0$, $B$ and $\widetilde{h}_1$ are fixed. Among these parameters, only $\gamma_{TX,1}$, $\gamma_{RX,1}$ and $\overline{\gamma_{RX,1}}$ are associated with $\mathbf{w}_B$ and $\mathbf{w}_M$, which vary in the beam alignment procedure. Note that in (\ref{Sum_J}), $\mathbf{J}_{\sum_{}^{}}$ can be equivalently written as another form:
\begin{equation}\label{Sum_J_2}
\mathbf{J}_{\sum_{}^{}}=M\times\mathbb{E}_{(\mathbf{w}_B,\mathbf{w}_M)}\left[\mathbf{J}_\mathbf{g}(\mathbf{w}_B,\widetilde{\mathbf{\Theta}}_1,\mathbf{w}_M)\right]
=\varpi\times\frac{T_c}{T_s}\times\mathbf{T}\mathbf{A}\mathbf{T}^T
\end{equation}
where $\mathbf{A}=\mathbb{E}_{(\mathbf{w}_B,\mathbf{w}_M)}\left[\mathbf{J}_{\bm{\eta}}(\mathbf{w}_B,\widetilde{\mathbf{\Theta}}_1,\mathbf{w}_M)\right]$.

To derive the closed-form expression, we should calculate $\mathbf{A}$, which is equivalent to calculating $\mathbb{E}_{(\mathbf{w}_B,\mathbf{w}_M)}\left[\gamma_{TX,1}\right]$, $\mathbb{E}_{(\mathbf{w}_B,\mathbf{w}_M)}\left[\gamma_{RX,1}\right]$, $\mathbb{E}_{(\mathbf{w}_B,\mathbf{w}_M)}\left[\overline{\gamma_{RX,1}}\right]$, $\mathbb{E}_{(\mathbf{w}_B,\mathbf{w}_M)}\left[|\gamma_{TX,1}|^2\right]$, $\mathbb{E}_{(\mathbf{w}_B,\mathbf{w}_M)}\left[|\gamma_{RX,1}|^2\right]$, $\mathbb{E}_{(\mathbf{w}_B,\mathbf{w}_M)}\left[|\overline{\gamma_{RX,1}}|^2\right]$ and $\mathbb{E}_{(\mathbf{w}_B,\mathbf{w}_M)}\left[\overline{\gamma_{RX,1}}^*\gamma_{RX,1}\right]$ according to (\ref{2-J_tau1_tau1}) to (\ref{66-J_hR1_hI1}).

After a few manipulations, we obtain the expressions of the elements in $\mathbf{A}$, denoted by $A_{i,j}$ for $i=1,2...,6$ and $j=1,2,...,6$, in Appendix C, and show that \emph{$\mathbf{A}$ is independent of or does not change with $\varpi$}. As a result, we have
\begin{equation}\label{PEB_Closedform}
PEB(\varpi)=\sqrt{tr\left\{\left[\mathbf{J}_{\sum_{}^{}}^{-1}\right]_{1:2,1:2}\right\}}=\frac{1}{\sqrt{\varpi}}\times\sqrt{\frac{T_s}{T_c}tr\left\{\left[\left(\mathbf{T}\mathbf{A}\mathbf{T}^T\right)^{-1}\right]_{1:2,1:2}\right\}}
\end{equation}
\begin{equation}\label{REB_Closedform}
REB(\varpi)=\sqrt{tr\left\{\left[\mathbf{J}_{\sum_{}^{}}^{-1}\right]_{3,3}\right\}}=\frac{1}{\sqrt{\varpi}}\times\sqrt{\frac{T_s}{T_c}tr\left\{\left[\left(\mathbf{T}\mathbf{A}\mathbf{T}^T\right)^{-1}\right]_{3,3}\right\}}
\end{equation}
which are inversely proportional to $\sqrt{\varpi}$.

\subsection{EADR}

\subsubsection{Mathematical Description}

$\ $

According to (\ref{received effective signal}) and the time allocation pattern in Figure \ref{Time}, the EADR is expressed as
\begin{equation}\label{EADR}
\begin{split}
R_{eff}=\left(1-\frac{MT_s+T_o}{T_c}\right)B\times\log_2\left(1+\frac{P_{TX}|\widetilde{h}_1|^2|\widetilde{\mathbf{w}}_M^H\mathbf{H}_{IM} \widetilde{\mathbf{\Theta}}_2 \mathbf{H}_{BI}\widetilde{\mathbf{w}}_B|^2}{N_0B}\right)
\end{split}
\end{equation}
%where $\widetilde{\mathbf{w}}_M \in\bm{\mathcal{C}}_{MU}$ and $\widetilde{\mathbf{w}}_B \in\bm{\mathcal{C}}_{BS}$ satisfy
%\begin{equation}\label{optimal beamformer pair}
%(\widetilde{\mathbf{w}}_M,\widetilde{\mathbf{w}}_B)=\mathop{\arg\max}_{(\mathbf{w}_M,\mathbf{w}_B)}
%\left(\frac{P_{TX}|\widetilde{h}_1|^2|\mathbf{w}_M^H\mathbf{H}_{IM} \widetilde{\mathbf{\Theta}}_1 \mathbf{H}_{BI}\mathbf{w}_B|^2}{N_0B}\right)
%\end{equation}

$\ $

\subsubsection{Phase Shift Design}

$\ $

The optimal configuration of $\widetilde{\mathbf{\Theta}}_2$ in the $l$-th communication period should make the EADR achieve its maximum. According to (\ref{EADR}), we have 
\begin{equation}\label{SNR_Fenzi}
|\widetilde{\mathbf{w}}_M^H\mathbf{H}_{IM} \widetilde{\mathbf{\Theta}}_2 \mathbf{H}_{BI}\widetilde{\mathbf{w}}_B|^2\leq
N_M N^2 N_B
\end{equation}

For simple analysis, we assume that the antenna arrays have high angular resolution when $N_B$ and $N_M$ is large, so that $\widetilde{\mathbf{w}}_M$ and $\widetilde{\mathbf{w}}_B$ approximately satisfy $\widetilde{\mathbf{w}}_M\approx\frac{1}{\sqrt{N_M}}\mathbf{a}_{RX}(\varphi_{RX,1})$ and $\widetilde{\mathbf{w}}_B\approx\frac{1}{\sqrt{N_B}}\mathbf{a}_{TX}(\varphi_{TX,1})$. Therefore, we have $|\widetilde{\mathbf{w}}_M^H\mathbf{H}_{IM} \widetilde{\mathbf{\Theta}}_2 \mathbf{H}_{BI}\widetilde{\mathbf{w}}_B|^2\approx
N_M N^2 N_B$ when 
$\widetilde{\mathbf{\Theta}}_2=diag\left(\mathbf{a}_{IRS}(\varphi_{IRS,2}^a,\varphi_{IRS,2}^e)\odot\mathbf{a}_{IRS}^*(\varphi_{IRS,1}^a,\varphi_{IRS,1}^e)\right)$,
%with $\mathbf{k}_M$ and $\mathbf{k}_B$ given by
%\begin{equation}
%\mathbf{k}_M=\sqrt{N_M}\left[\widetilde{\mathbf{w}}_M^H \mathbf{a}_{RX}(\varphi_{RX,1})\right]^{-1}\mathbf{1}^T
%\in\mathbb{C}^{N\times 1}
%\end{equation}
%\begin{equation}
%\mathbf{k}_B=\sqrt{N_B}\left[\mathbf{a}_{TX}^H(\varphi_{TX,1}) \widetilde{\mathbf{w}}_B\right]^{-1}\mathbf{1}^T
%\in\mathbb{C}^{N\times 1}
%\end{equation}
Because the IRS phase shifts are adjusted based on the estimated parameters, the optimal $\widetilde{\mathbf{\Theta}}_2$ is designed as
\begin{equation}\label{Optimal Theta_2}
\widetilde{\mathbf{\Theta}}_2=
diag\left(\mathbf{a}_{IRS}(\widehat{\varphi_{IRS,2}^a}(l),\widehat{\varphi_{IRS,2}^e}(l))\odot\mathbf{a}_{IRS}^*(\varphi_{IRS,1}^a,\varphi_{IRS,1}^e)\right)
\end{equation}
%where $\widehat{\mathbf{k}_M}=\sqrt{N_M}\left[\widetilde{\mathbf{w}}_M^H \mathbf{a}_{RX}(\widehat{\varphi_{RX,1}})\right]^{-1}\mathbf{1}^T$.
where $\widehat{\varphi_{IRS,2}^a}(l)$ and $\widehat{\varphi_{IRS,2}^e}(l)$, according to Figure \ref{Working Process}, stand for the estimated $\varphi_{IRS,2}^a$ and $\varphi_{IRS,2}^e$ in the $l$-th communication period.

$\ $

\subsubsection{Approximate Closed-form Expression}

$\ $

Here, if the estimation errors are assumed to be slight, i.e. $\widehat{x}\approx x$ for variable $x$, by substituting (\ref{Optimal Theta_2}) into (\ref{EADR}), we obtain
\begin{equation}\label{EADR_max}
\begin{split}
R_{eff}(\varpi)\approx&\left(1-\frac{MT_s+T_o}{T_c}\right)B\times\log_2\left(1+\frac{P_{TX}|\widetilde{h}_1|^2N_M N^2 N_B}{N_0B}\right)\\
=&\left(1-\frac{T_o}{T_c}-\varpi\right)B\times \log_2\left(1+\frac{P_{TX}|\widetilde{h}_1|^2 N^2 T_c}{N_0BT_s}\varpi\right)
\end{split}
\end{equation}
which is an approximate function of $\varpi$ when the other parameters are given.

%Eq. (\ref{effective data rate}) demonstrates that the effective data rate is primarily related to $(\widetilde{\mathbf{w}}_M,\widetilde{\mathbf{w}}_B)$ and the number of the searched beamforming vectors ($M=N_B\times N_M$, which is also equivalent to the product of the codebook sizes or the product of the transmitting and receiving antenna sizes). It is remarkable that: 1) larger codebook sizes causes more FIM under the exhaustive search strategy, thus reduces the PEB/REB; 2) expanding the codebook sizes results in an improvement of the received SNR to some extent, but leads to an increase of the number of the searched beamforming vectors, which prolongs the beam alignment procedure and then shortens the effective data transmission period. Therefore, there exists a trade-off between the position/rotation estimation accuracies and the effective communication data rate. In the next section, we will provide a joint optimization approach on both the PEB/REB and the effective communication data rate.

\section{Trade-off and Joint Optimization}

This section discusses the trade-off between PEB/REB and EADR, and proposes an algorithm to find the joint optimal solution for the potential localization and data-transmission performances by optimizing the time allocation ratio $\varpi$.

\subsection{Trade-off between PEB/REB and EADR}

From (\ref{PEB_Closedform}), (\ref{REB_Closedform}) and (\ref{EADR_max}), it is indicated that as $\varpi$ grows, the PEB and REB continuously decreases, resulting in an improvement of the potential localization performance, while the EADR varies in a non-monotonic way. The occurrence of this phenomenon can further be explicated by retrospecting the system working process in Section II. As shown in Figure \ref{Time} and \ref{Working Process} in Section II, when $T_b$ is extended, more beams are searched and more pilot signals are transmitted during the BALS, leading to a higher positioning accuracy. Meanwhile, with more beams searched, the codebook size at the BS is expanded and more antennas are concomitantly activated, leading to a higher received SNR. However, as $T_c$ and $T_o$ are fixed, prolonging the BALS shortens the EDTS and reduces $\left(1-\frac{T_o}{T_c}-\varpi\right)$, which, according to (\ref{EADR_max}), influences the EADR dominantly. Therefore, there exists a trade-off between PEB/REB and EADR. In view of this trade-off, we will jointly optimize the two performance metrics in the remainder of this section.

\subsection{Joint Optimization}

%Moreover, because according to Figure \ref{Working Process}, the performance metric is practically calculated using estimated parameters, we also denote the estimated PEB and REB as $\widehat{PEB}$ and $\widehat{PEB}$, which we obtain by replacing 

Before the joint optimization problem is formulated, an objective function as a weighted sum of (PEB + REB) and EADR with respect to $\varpi$ is first constructed. Since the system can only acquire the estimates of the position/orientation and channel parameters related to the MU, based on (\ref{PEB_Closedform}), (\ref{REB_Closedform}) and (\ref{EADR_max}), we define
\begin{equation}\label{closed-form PREB}
\widehat{PREB}(\varpi)=\widehat{PEB}(\varpi)+\widehat{REB}(\varpi)=\frac{1}{\sqrt{\varpi}}\widehat{\mathfrak{X}}
\end{equation}
\begin{equation}\label{closed-form EADR}
\widehat{R_{eff}}(\varpi)\approx B\left(1-\frac{T_o}{T_c}-\varpi\right)\log_2\left(1+\widehat{\mathfrak{Y}}\varpi\right)
\end{equation}
with $\widehat{\mathfrak{X}}$ and $\widehat{\mathfrak{Y}}$ given by
\begin{equation}
\widehat{\mathfrak{X}}=\sqrt{\frac{T_s}{T_c}tr\left(\left[(\widehat{\mathbf{T}}\widehat{\mathbf{A}}\widehat{\mathbf{T}}^T)^{-1}\right]_{1:2,1:2}\right)}
+\sqrt{\frac{T_s}{T_c}tr\left(\left[(\widehat{\mathbf{T}}\widehat{\mathbf{A}}\widehat{\mathbf{T}}^T)^{-1}\right]_{3,3}\right)}
\end{equation}
\begin{equation}
\widehat{\mathfrak{Y}}=\frac{P_{TX}|\widehat{\widetilde{h}_1}|^2 N^2 T_c}{N_0BT_s}
\end{equation}
where $\widehat{\mathbf{T}}$, $\widehat{\mathbf{A}}$ and $\widehat{\widetilde{h}_1}$ are the estimates of $\mathbf{T}$, $\mathbf{A}$ and $\widetilde{h}_1$. Specifically, $\widehat{\mathbf{T}}$ and $\widehat{\mathbf{A}}$ are obtained by replacing $\mathbf{p}$, $\alpha$, $\widetilde{h}_1$ and the corresponding $\varphi_{RX,1}$, $\varphi_{IRS,2}^a$, $\varphi_{IRS,2}^e$ in $\mathbf{T}$ and $\mathbf{A}$ with $\widehat{\mathbf{p}}$, $\widehat{\alpha}$, $\widehat{\widetilde{h}_1}$ and the corresponding $\widehat{\varphi_{RX,1}}$, $\widehat{\varphi_{IRS,2}^a}$, $\widehat{\varphi_{IRS,2}^e}$.
Then, the joint optimization problem is formulated as
\begin{subequations}
\begin{align}
(P1):\ \ \min_{\varpi>0}\ \  &{\widehat{PREB}(\varpi)-\xi \widehat{R_{eff}}(\varpi)}\\
s.t.\ \  & \varpi-\left(1-\frac{T_o}{T_c}\right)\leq0
\end{align}
\end{subequations}
where $\xi$ represents a predetermined weight parameter.
Subsequently, in order to solve $(P1)$, we construct a Lagrangian function by introducing a multiplier $\lambda_1$:
\begin{equation}
L(\varpi,\lambda_1)=\widehat{PREB}(\varpi)-\xi \widehat{R_{eff}}(\varpi)+\lambda_1\left[\varpi-\left(1-\frac{T_o}{T_c}\right)\right]
\end{equation}

According to the KKT conditions, the optimal $\varpi$ should satisfy:

\begin{subequations}\label{KKT conditions}
\begin{align}
\frac{\partial L(\varpi,\lambda_1)}{\partial\varpi}=
-\frac{1}{2}\varpi^{-\frac{3}{2}}\widehat{\mathfrak{X}}-&\xi\left[\frac{\widehat{\mathfrak{Y}}B\left(1-\frac{T_o}{T_c}-\varpi\right)}{(1+\widehat{\mathfrak{Y}}\varpi)\ln2} - B\log_2\left(1+\widehat{\mathfrak{Y}}\varpi\right)\right]+\lambda_1=0\\
&\lambda_1\left[\varpi-\left(1-\frac{T_o}{T_c}\right)\right]=0\\
&\lambda_1\geq0\\
&0<\varpi\leq1-\frac{T_o}{T_c}
\end{align}
\end{subequations}

From (\ref{KKT conditions}a) to (\ref{KKT conditions}d), it is noted that: 1) if $\lambda_1=0$, we need to solve
\begin{equation}\label{Eqset_1}
-\frac{1}{2}\varpi^{-\frac{3}{2}}\widehat{\mathfrak{X}}-\xi\left[\frac{\widehat{\mathfrak{Y}}B\left(1-\frac{T_o}{T_c}-\varpi\right)}{(1+\widehat{\mathfrak{Y}}\varpi)\ln2} - B\log_2\left(1+\widehat{\mathfrak{Y}}\varpi\right)\right]=0
\end{equation}
and obtain $\varpi=\widetilde{\varpi}_1$. If $\widetilde{\varpi}_1$ satisfies $0<\widetilde{\varpi}_1\leq 1-\frac{T_o}{T_c}$, it is a solution which meets the KKT conditions. 2) If $\lambda_1\neq0$, we need to solve
\begin{subequations}\label{Eqset_2}
\begin{align}
-\frac{1}{2}\varpi^{-\frac{3}{2}}\widehat{\mathfrak{X}}-\xi&\left[\frac{\widehat{\mathfrak{Y}}B\left(1-\frac{T_o}{T_c}-\varpi\right)}{(1+\widehat{\mathfrak{Y}}\varpi)\ln2} - B\log_2\left(1+\widehat{\mathfrak{Y}}\varpi\right)\right]+\lambda_1=0\\
&\varpi-\left(1-\frac{T_o}{T_c}\right)=0
\end{align}
\end{subequations}
and obtain $\varpi=\widetilde{\varpi}_2=1-\frac{T_o}{T_c}$ and $\lambda_1=\frac{1}{2}\left(1-\frac{T_o}{T_c}\right)^{-\frac{3}{2}}\widehat{\mathfrak{X}}-\xi B\log_2\left[1+\widehat{\mathfrak{Y}}\left(1-\frac{T_o}{T_c}\right)\right]$. If $\frac{1}{2}\left(1-\frac{T_o}{T_c}\right)^{-\frac{3}{2}}\widehat{\mathfrak{X}}-\xi B\log_2\left[1+\widehat{\mathfrak{Y}}\left(1-\frac{T_o}{T_c}\right)\right]\geq0$, $\widetilde{\varpi}_2$ is also a solution that meets the KKT conditions. Finally, if $\widetilde{\varpi}_1$ and $\widetilde{\varpi}_2$ both satisfy the KKT conditions, the optimal $\varpi$, denoted by $\widetilde{\varpi}$, is obtained by
\begin{equation}\label{Optimal_varpi}
\widetilde{\varpi}=\mathop{\arg\min}_{\varpi=\widetilde{\varpi}_1,\widetilde{\varpi}_2}\left\{\widehat{PREB}(\varpi)-\xi \widehat{R_{eff}}(\varpi)\right\}
\end{equation}

Based on the above analysis, we design the \textbf{Algorithm 1} to minimize the objective function in (P1) and find the optimal solution of $\varpi$. Forasmuch as the performance of \textbf{Algorithm 1} is closely related to the estimation uncertainty, the influences of the estimation errors on the optimization performance will be discussed in the following Section V.

\begin{algorithm}
\caption{Joint Optimization Algorithm for Solving (P1)}
\LinesNumbered
{\bf Input:} The estimated parameters including $\widehat{\widetilde{h}_1}$, $\widehat{\mathbf{p}}$, $\widehat{\alpha}$, the corresponding AOAs and AODs\;
Compute $\widehat{\mathfrak{X}}$ and $\widehat{\mathfrak{Y}}$, initialize $\xi$\;
Obtain $\varpi=\widetilde{\varpi}_1$ by solving (\ref{Eqset_1})\;
Set $\widetilde{\varpi}_2=1-\frac{T_o}{T_c}$ and calculate $\lambda_1=\frac{1}{2}\left(1-\frac{T_o}{T_c}\right)^{-\frac{3}{2}}\widehat{\mathfrak{X}}-\xi B\log_2\left[1+\widehat{\mathfrak{Y}}\left(1-\frac{T_o}{T_c}\right)\right]$\;
     \If{$0<\widetilde{\varpi}_1\leq 1-\frac{T_o}{T_c}$}{
     \If{$\lambda_1\geq0$}{
        Obtain the optimal $\varpi$ by calculating $\widetilde{\varpi}=\mathop{\arg\min}_{\varpi=\widetilde{\varpi}_1,\widetilde{\varpi}_2}\left\{\widehat{PREB}(\varpi)-\xi \widehat{R_{eff}}(\varpi)\right\}$\;
        \Else{Obtain the optimal $\varpi$ from $\widetilde{\varpi}=\widetilde{\varpi}_1$;}
      }
      \ElseIf{$\lambda_1\geq0$}{
      Obtain the optimal $\varpi$ from $\widetilde{\varpi}=\widetilde{\varpi}_2=1-\frac{T_o}{T_c}$\;
      \Else{Empty solution for optimal $\varpi$;}
      }
      }
{\bf Output:} $\widetilde{\varpi}$ as the optimal time allocation ratio;

\end{algorithm}

\section{Numerical Results}
This section presents the simulation results of the performance metrics as well as their trade-off, and investigates the joint optimization performance of the proposed algorithm, in the presence of different levels of user mobility and estimation uncertainty.

\subsection{System Parameters}
Before the simulations, we set the system parameters in Table I by referring to \cite{Shahmansoori2018(TWC), Destino2017(ICC)}. According to the parameters, we further obtain the signal wavelength $\lambda=c/f_c\approx 5$ mm, the antenna spacing $d=\lambda/2=2.5$ mm, and the distances between BS and MU ($d_0=60$ m), BS and IRS ($d_{1,1}=30$ m) and IRS and MU ($d_{1,2}\approx 53.85$ m).

\begin{table}
\renewcommand{\arraystretch}{1.3} 
\caption{Parameter configuration of the IMM-JLCS.}
\label{Table_Parameters}
\centering
\begin{small} 
\begin{tabular}{ccc}
\hline
Parameters & Definitions & Values\\
\hline
Position Coordinate of the BS & $(q_x,q_y,\beta_{BS})$ & $(0,0,40)$ (m)\\
Position Coordinate of the IRS & $(v_x,v_y,\beta_{IRS})$ & $(-20,20,30)$ (m)\\
Position Coordinate of the MU & $(p_x,p_y,0)$ & $(20,40,0)$ (m)\\
Rotation Angle of the MU & $\alpha$ & $\pi/4$ (rad)\\
Transmit Power & $P_{TX}$ & $27$ (dBm)\\
Noise Power & $\sigma_w^2$ & $-80$ (dBm)\\
Number of Antennas on BS/MU & $N_B^t=N_M^r$ & $32$\\
Carrier Frequency & $f_c$ & $60$ (GHz)\\
Signal Bandwidth & $B$ & $100$ (MHz)\\
Reflection Coefficient & $\delta$ & 1 \\
Power Attenuation Coefficient & $\zeta$ & 1 \\
Complex Channel Coefficient & $h_1$ & $e^{j2\pi\times rand(0,1)}$\\
Duration of the Pilot Signal & $T_s$ & $67$ (us)\\
Duration of the Joint Optimization Procedure & $T_o$ & $1$ (ms)\\
Duration of the Entire Communication Period & $T_c$ & $N_B^t\times N_M^r\times T_s + T_o=69.608$ (ms)\\
\hline
\end{tabular}
\end{small}
\end{table} 

\subsection{Performance Metrics and Trade-off}

First, we numerically investigate the trade-off between PEB/REB and EADR, and discuss the impact of the user mobility and localization error on the performances. 

\begin{figure*}[!t]
\centering
\subfloat[]{\includegraphics[width=3.2in]{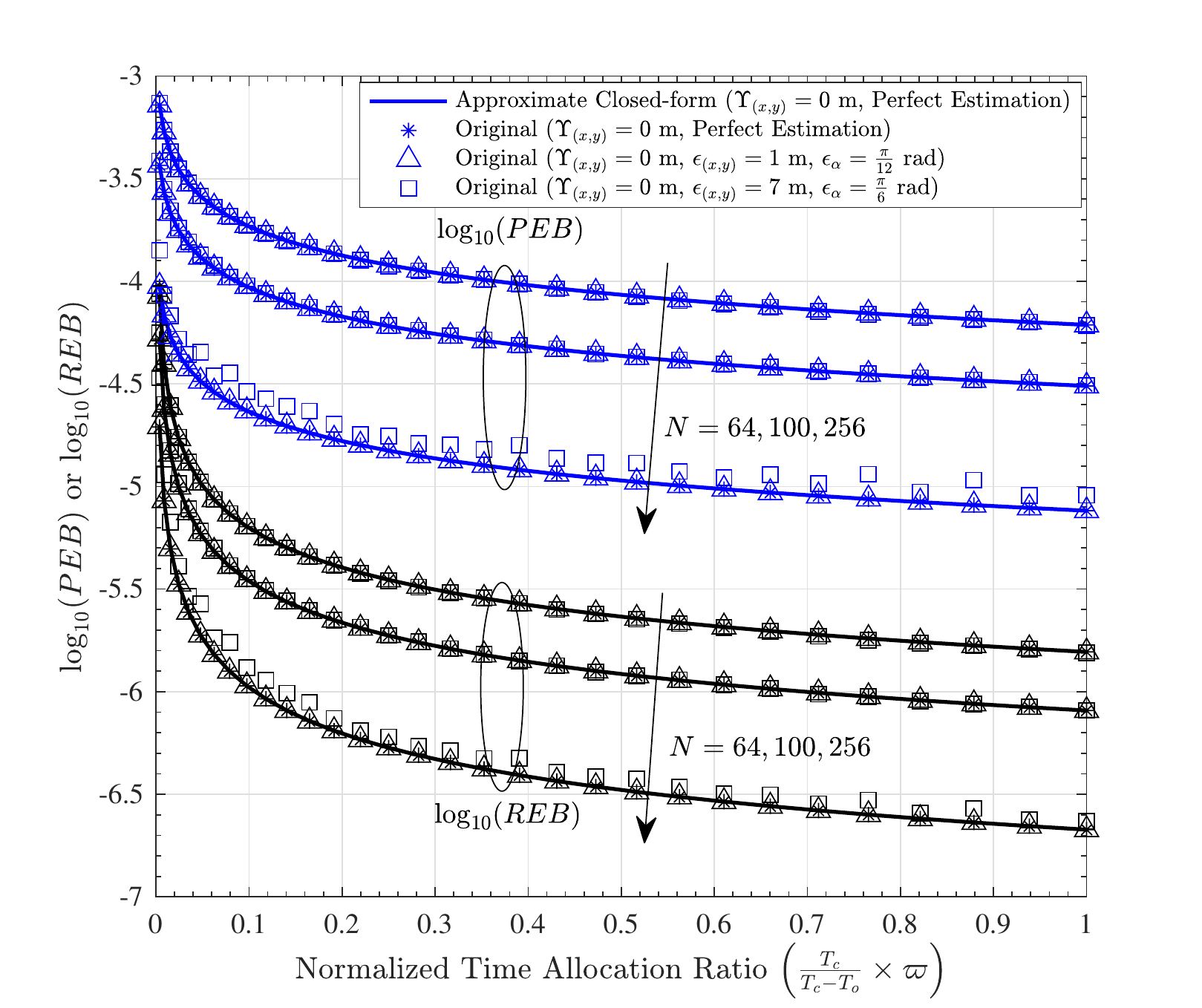}}
\label{PEB_and_REB_Time_Imperfect}
\subfloat[]{\includegraphics[width=3.2in]{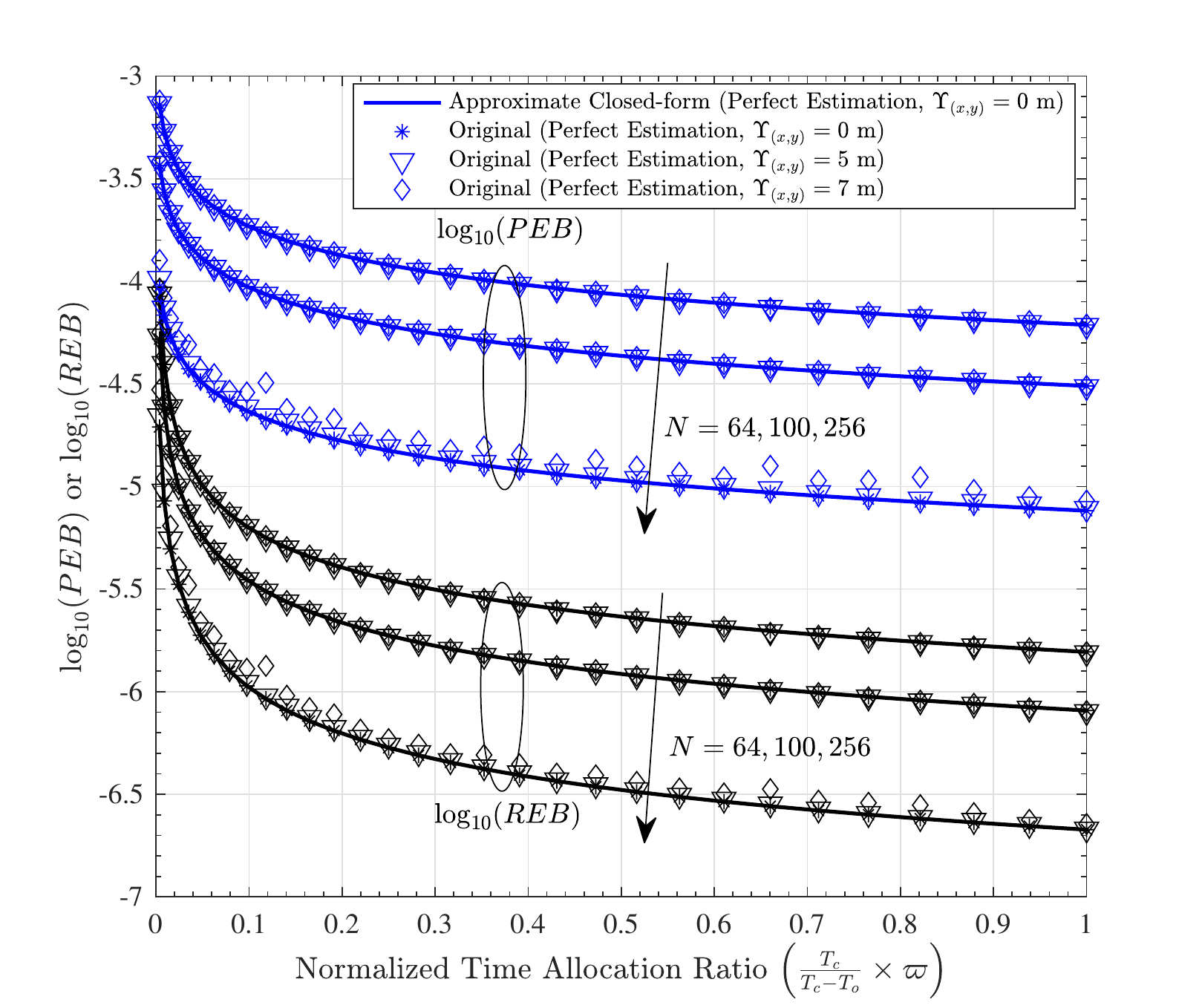}}
\label{PEB_and_REB_Time_Movement}
\hfil
\caption{$\log_{10}(PEB)$ and $\log_{10}(REB)$ as functions of $\left(\frac{T_c}{T_c-T_0}\times\varpi\right)$, with: (a) different $\epsilon_{(x,y)}$, $\epsilon_{\alpha}$ and $N$; (b) different $\Upsilon_{(x,y)}$ and $N$. The curves with legend "Approximate Closed-form" represent the results of (\ref{PEB_Closedform}) and (\ref{REB_Closedform}), while the marks with legend "Original" represent the results of (\ref{PEB}) and (\ref{REB}). The discrete marks represent the results averaged on 50 Monte Carlo trials.}
\label{PEB_and_REB_Time}
\end{figure*}

Figure \ref{PEB_and_REB_Time} displays the $\log_{10}(PEB)$ and $\log_{10}(REB)$ as functions of the normalized time allocation ratio $\left(\frac{T_c}{T_c-T_0}\times\varpi\right)$, in the presence of different levels of (a) localization error, and (b) user mobility.
Both Figure \ref{PEB_and_REB_Time} (a) and Figure \ref{PEB_and_REB_Time} (b) indicate that: 1) as $\varpi$ or $N$ grows, $\log_{10}(PEB)$ and $\log_{10}(REB)$ decrease, leading to a better potential position/orientation estimation performance. 2) The localization error with $\epsilon_{(x,y)}\leq 7$ m and $\epsilon_{\alpha}\leq \frac{\pi}{6}$ rad, and user mobility with $\Upsilon_{(x,y)}\leq 7$ m, slightly impact (increase) $\log_{10}(PEB)$ and $\log_{10}(REB)$ because they influence the IRS phase shift design for $\widetilde{\mathbf{\Theta}}_1$. 3) Under the assumption of perfect estimation, i.e. no estimation error, the approximate closed-form expressions in (\ref{PEB_Closedform}) and (\ref{REB_Closedform}) coincide with the original (\ref{PEB}) and (\ref{REB}), testifying the correctness of the derivations in Appendix C.

\begin{figure}[!t]
\includegraphics[width=3.2in]{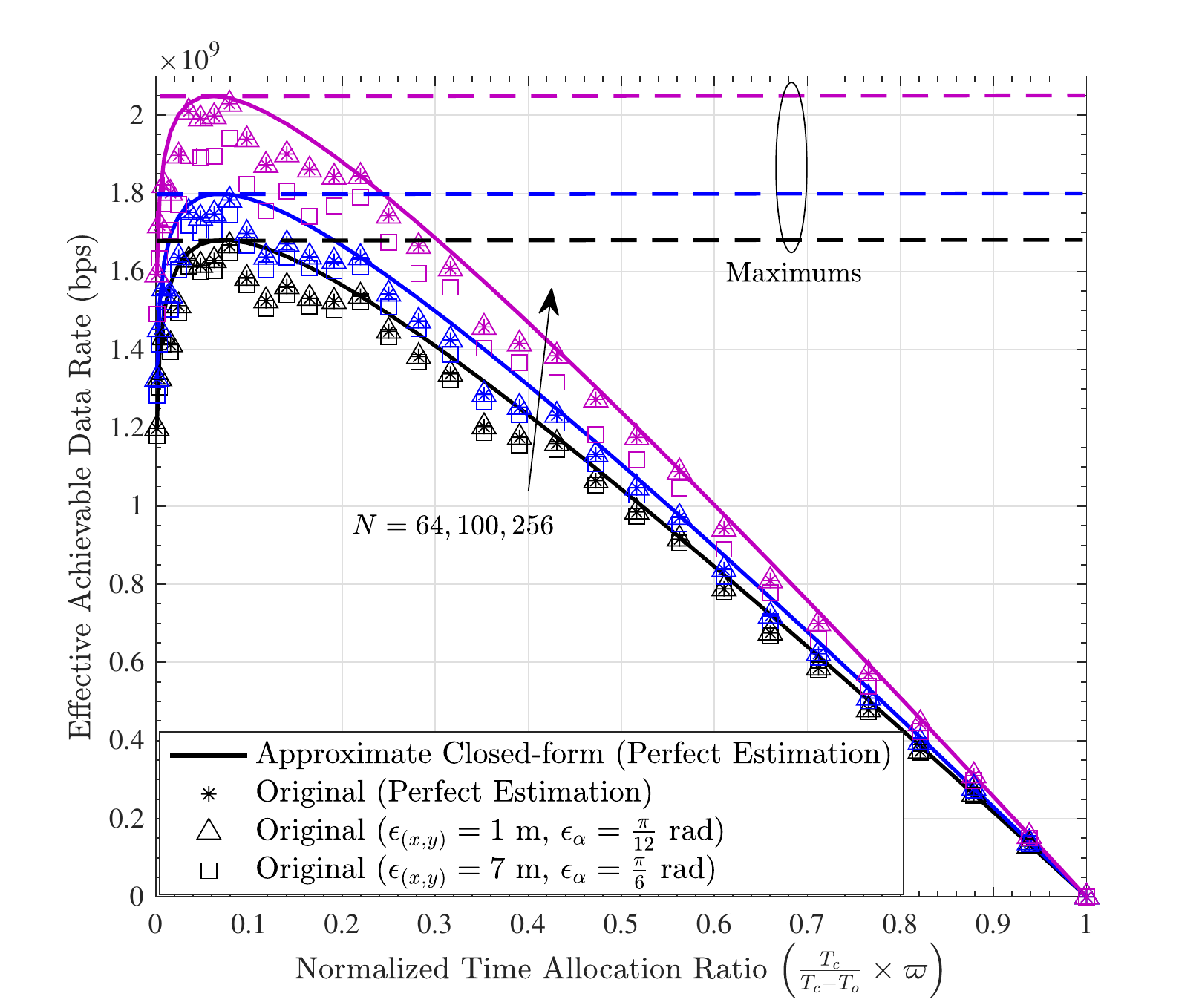}
\hfil
\centering
\caption{EADRs as functions of $\left(\frac{T_c}{T_c-T_0}\times\varpi\right)$, with different $\epsilon_{(x,y)}$, $\epsilon_{\alpha}$ and $N$. The curves with legend "Approximate Closed-form" represent the results of (\ref{EADR_max}), while the marks with legend "Original" represent the results of (\ref{EADR}). The discrete marks represent the results averaged on 50 Monte Carlo trials.}
\label{EADR_Time_Imperfect}
\end{figure}

Figure \ref{EADR_Time_Imperfect} depicts the EADRs as functions of $\left(\frac{T_c}{T_c-T_0}\times\varpi\right)$, in the presence of different levels of localization error. It is demonstrated that: 1) as $N$ grows, the EADR increases, while as $\varpi$ grows, the EADR first ascends to a maximum and then descends to zero. 2) The localization error with $\epsilon_{(x,y)}\leq 7$ m and $\epsilon_{\alpha}\leq \frac{\pi}{6}$ rad degrade the EADR, as they influence the IRS phase shift design for $\widetilde{\mathbf{\Theta}}_2$. 3) Under the assumption of perfect estimation, the original (\ref{EADR}) is lower than the approximate closed-form expression in (\ref{EADR_max}) at several points. This is because the beam pair of $(\widetilde{\mathbf{w}}_M,\widetilde{\mathbf{w}}_B)$ is obtained from the codebooks, which may not precisely equal to $\left(\frac{1}{\sqrt{N_M}}\mathbf{a}_{RX}(\varphi_{RX,1}),\frac{1}{\sqrt{N_B}}\mathbf{a}_{TX}(\varphi_{TX,1})\right)$.

\begin{figure*}[!t]
\centering
\subfloat[]{\includegraphics[width=3.2in]{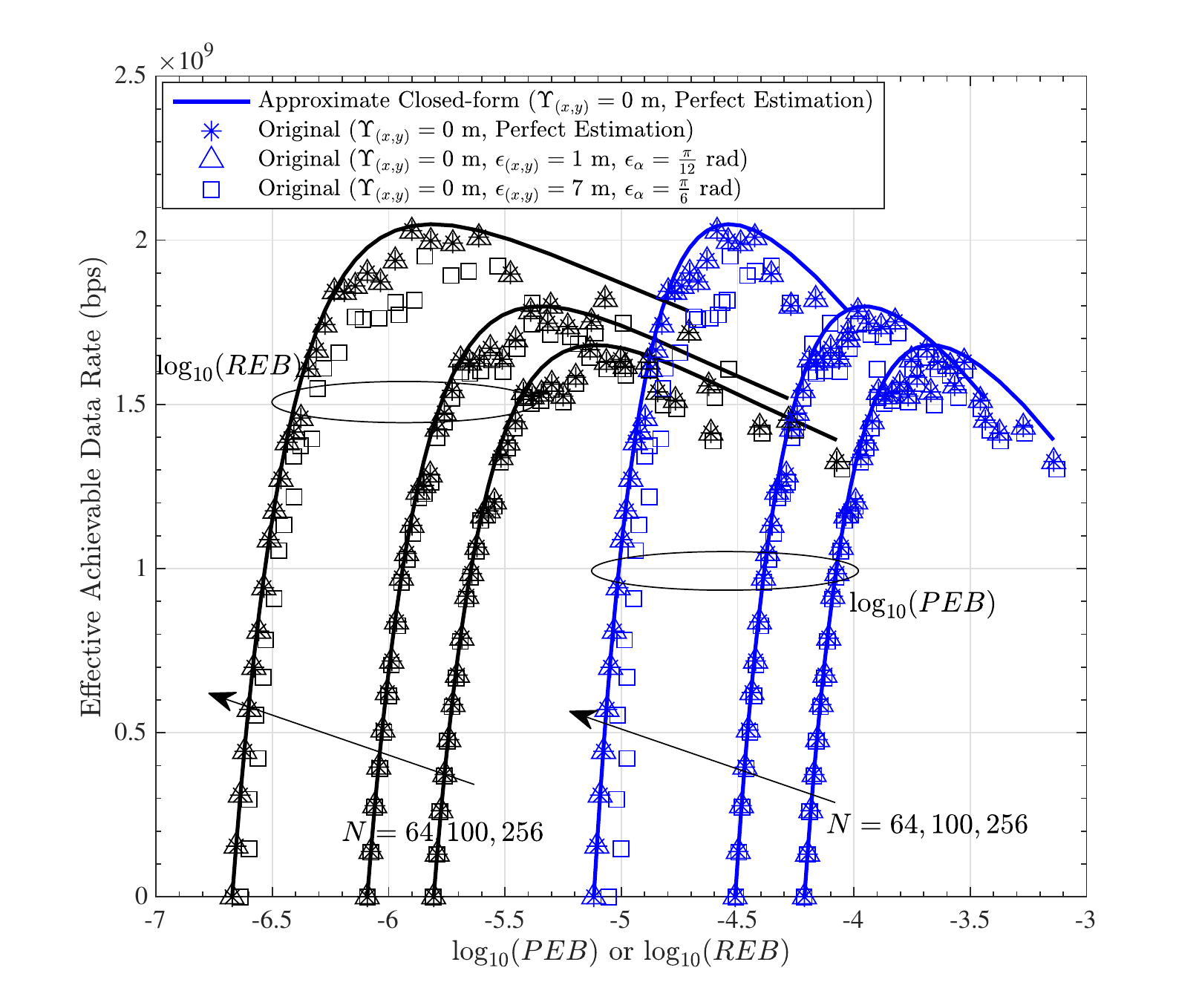}}
\label{PEB_and_REB_vs_EADR_Imperfect}
\subfloat[]{\includegraphics[width=3.2in]{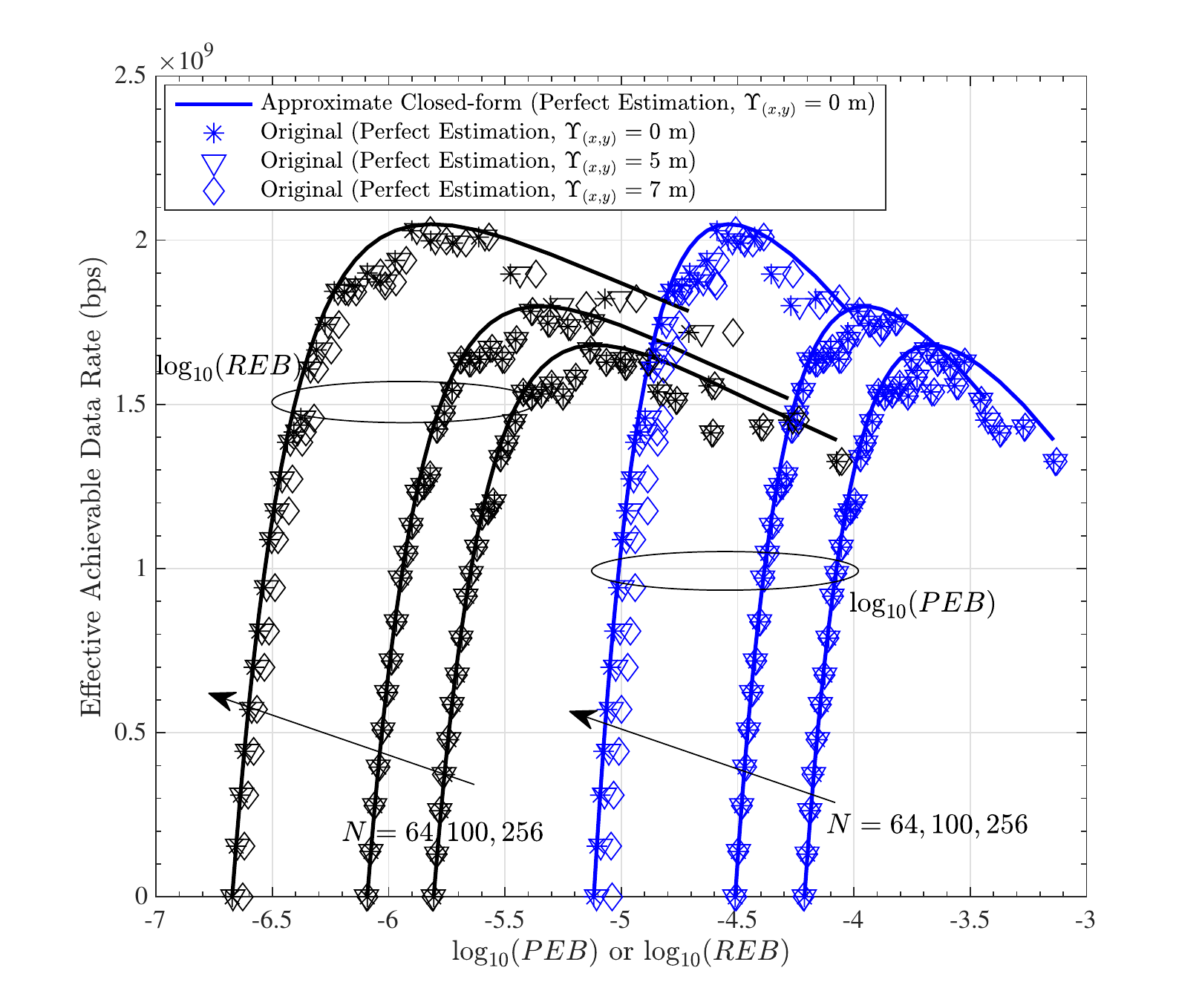}}
\label{PEB_and_REB_vs_EADR_Movement}
\hfil
\caption{EADRs as functions of $\log_{10}(PEB)$ and $\log_{10}(REB)$, with: (a) different $\epsilon_{(x,y)}$, $\epsilon_{\alpha}$ and $N$; (b) different $\Upsilon_{(x,y)}$ and $N$. The discrete marks represent the results averaged on 50 Monte Carlo trials.}
\label{PEB_and_REB_vs_EADR}
\end{figure*}

Figure \ref{PEB_and_REB_vs_EADR} plots the EADRs as functions of $\log_{10}(PEB)$ and $\log_{10}(REB)$, in the presence of different levels of (a) localization error, and (b) user mobility. It is illustrated that when the PEB/REB achieves the minimum, the EADR decreases to zero; when the EADR reaches its maximum, the PEB/REB does not achieve its own optimal state. Therefore, a trade-off exists between the PEB/REB and EADR, which share a joint optimal point represented by the peak of each curve.

%\begin{figure}[!t]
%\includegraphics[width=3.8in]{PEB_and_REB_Time_Imperfect-eps-converted-to.pdf}
%\hfil
%\centering
%\caption{$\log_{10}(PEB)$ and $\log_{10}(REB)$ as functions of $\left(\frac{T_c}{T_c-T_0}\times\varpi\right)$, with different $\epsilon_{(x,y)}$, $\epsilon_{\alpha}$ and $N$. The curves with legend "Approximate Closed-form" represent the results of (\ref{PEB_Closedform}) and (\ref{REB_Closedform}), while the marks with legend "Original" represent the results of (\ref{PEB}) and (\ref{REB}).}
%\label{PEB_and_REB_Time_Imperfect}
%\end{figure}

%\begin{figure}[!t]
%\includegraphics[width=3.8in]{PEB_and_REB_Time_Movement-eps-converted-to.pdf}
%\hfil
%\centering
%\caption{$\log_{10}(PEB)$ and $\log_{10}(REB)$ as functions of $\left(\frac{T_c}{T_c-T_0}\times\varpi\right)$, with different $\Upsilon_{(x,y)}$ and $N$. The curves with legend "Approximate Closed-form" represent the results of (\ref{PEB_Closedform}) and (\ref{REB_Closedform}), while the marks with legend "Original" represent the results of (\ref{PEB}) and (\ref{REB}).}
%\label{PEB_and_REB_Time_Movement}
%\end{figure}

%\begin{figure}[!t]
%\includegraphics[width=3.4in]{Error_Bound_versus_N-eps-converted-to.pdf}
%\hfil
%\caption{The PEB and REB as functions of the number of IRS reflecting elements $N$, with different number of antennas at the BS and the MU. The PEB and REB are measured by $log_{10}(XEB)$ for observation convenience, where $X$ is denoted by $X=P$ corresponding to PEB, or denoted by $X=R$ corresponding to REB.}
%\label{Error_Bound_versus_N}
%\end{figure}

\subsection{Comparisons with Random Phase Shifts}

The random IRS phase shifts can embody the reflection characteristic of scatterers without phase adjustment, which universally exist in the mmWave communication environment \cite{Shahmansoori2018(TWC)}. For evaluating the performance improvement brought by the IRS phase shift design, we compare $\widetilde{\mathbf{\Theta}}_1$ and $\widetilde{\mathbf{\Theta}}_2$ with random IRS phase shifts in terms of the PEB/REB and EADR performances.

\begin{figure*}[!t]
\centering
\subfloat[]{\includegraphics[width=2.1in]{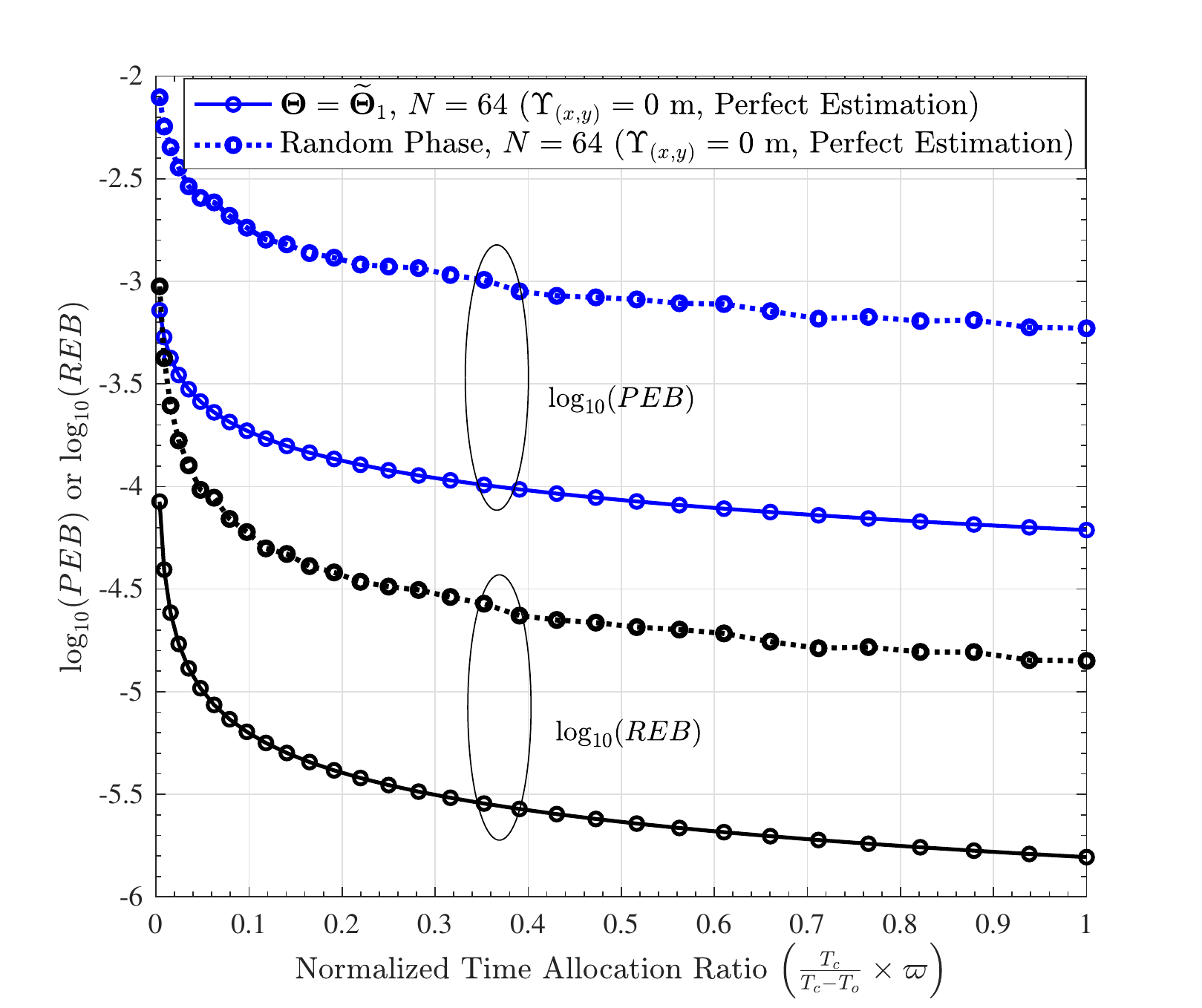}}
\label{Comparison_PEB_and_REB_Time}
\subfloat[]{\includegraphics[width=2.1in]{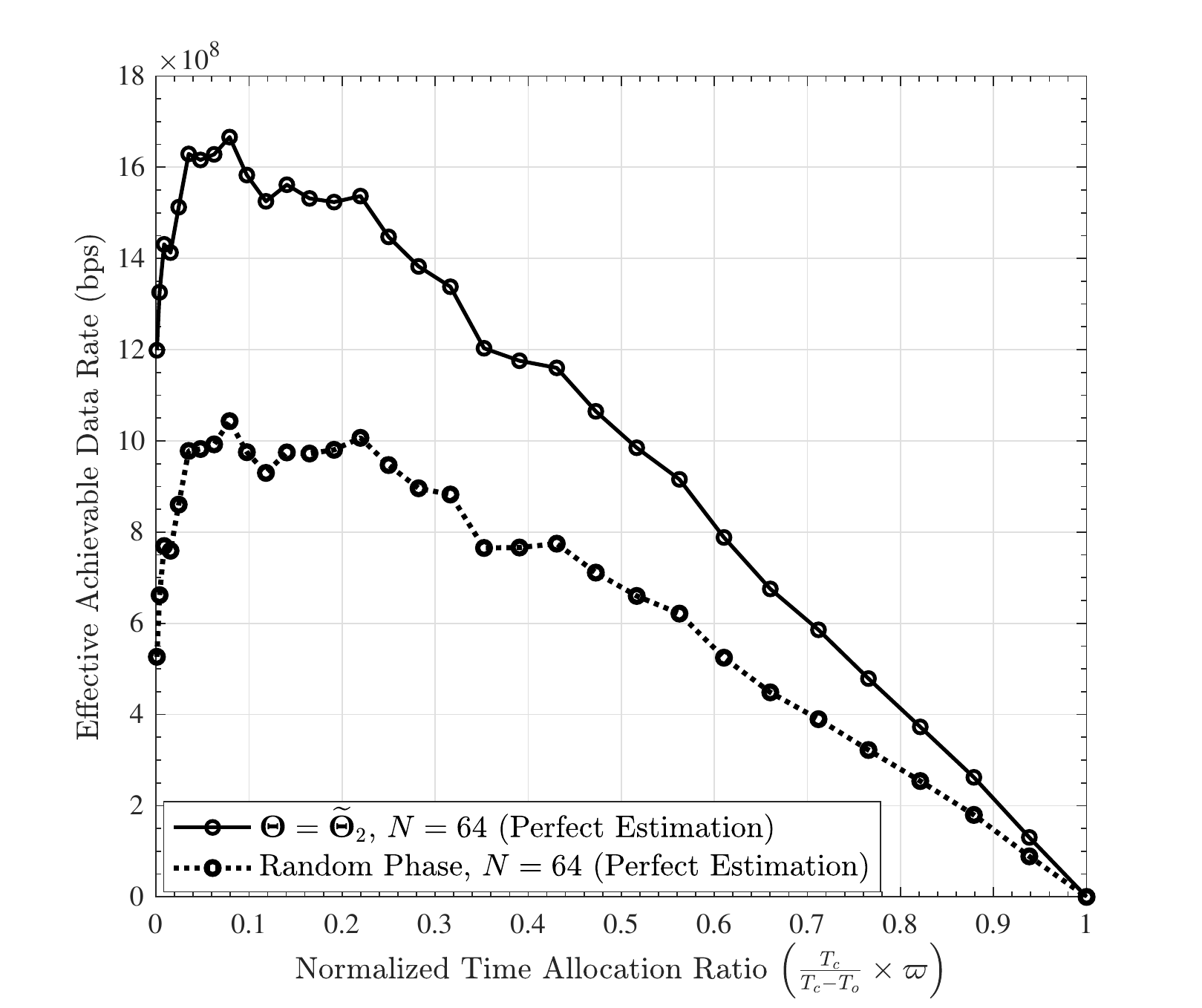}}
\label{Comparison_EADR_Time}
\subfloat[]{\includegraphics[width=2.1in]{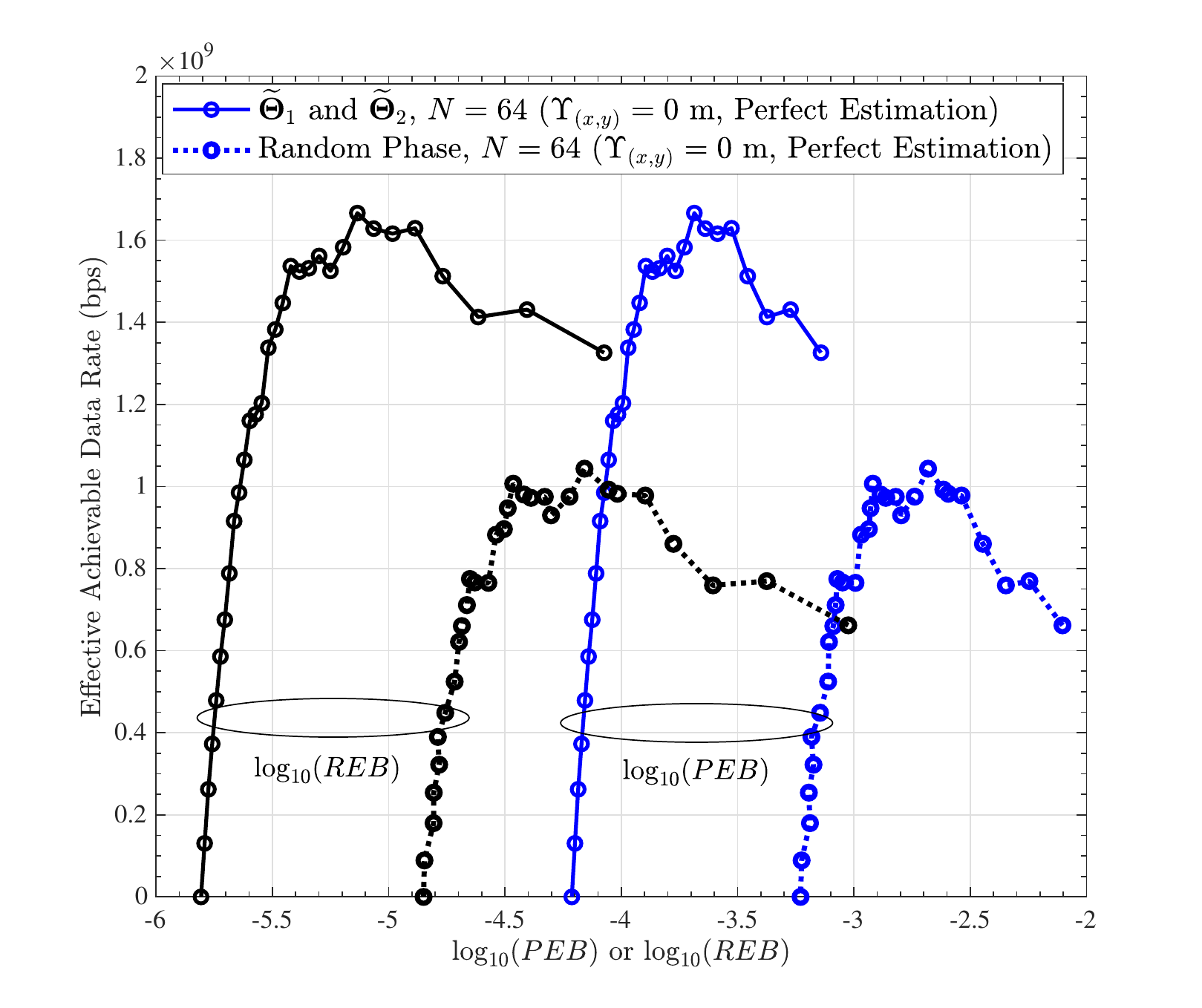}}
\label{Comparison_PEB_and_REB_vs_EADR}
\hfil
\caption{Comparisons with random IRS phase shifts when $N=64$ under the assumption perfect position/orientation estimation. The dotted curves represent the results averaged on 1000 Monte Carlo trials. (a) $\log_{10}(PEB)$ and $\log_{10}(REB)$ of (\ref{PEB}) and (\ref{REB}) as functions of $\left(\frac{T_c}{T_c-T_0}\times\varpi\right)$; (b) EADRs of (\ref{EADR}) as functions of $\left(\frac{T_c}{T_c-T_0}\times\varpi\right)$; (c) EADRs as functions of $\log_{10}(PEB)$ and $\log_{10}(REB)$.}
\label{Comparison_Random_Phase}
\end{figure*}

The comparisons are shown in Figure \ref{Comparison_Random_Phase}, where the random IRS phase shifts are uniformly distributed within $[-\pi,\pi]$. It is indicated that the PEB/REB with $\widetilde{\mathbf{\Theta}}_1$ and EADR with $\widetilde{\mathbf{\Theta}}_2$ are, respectively, lower and higher than those with random IRS phase shifts. This implies that an appropriate IRS phase shift configuration can improve the localization and data transmission performances to a large extent.

\subsection{Joint Optimization}

\begin{figure*}[!t]
\centering
\subfloat[]{\includegraphics[width=3.2in]{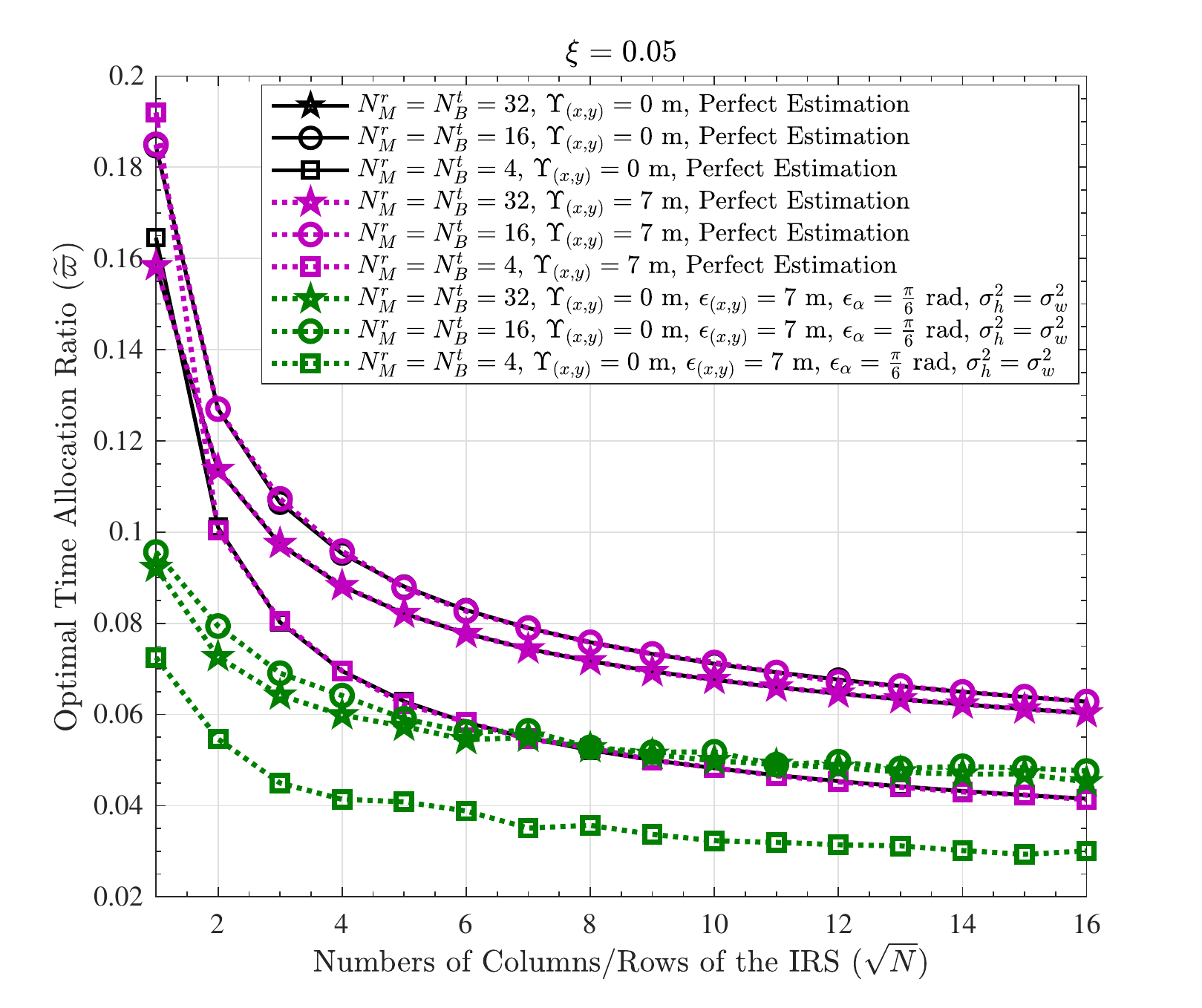}}
\label{Optimal_Varpi_L}
\subfloat[]{\includegraphics[width=3.2in]{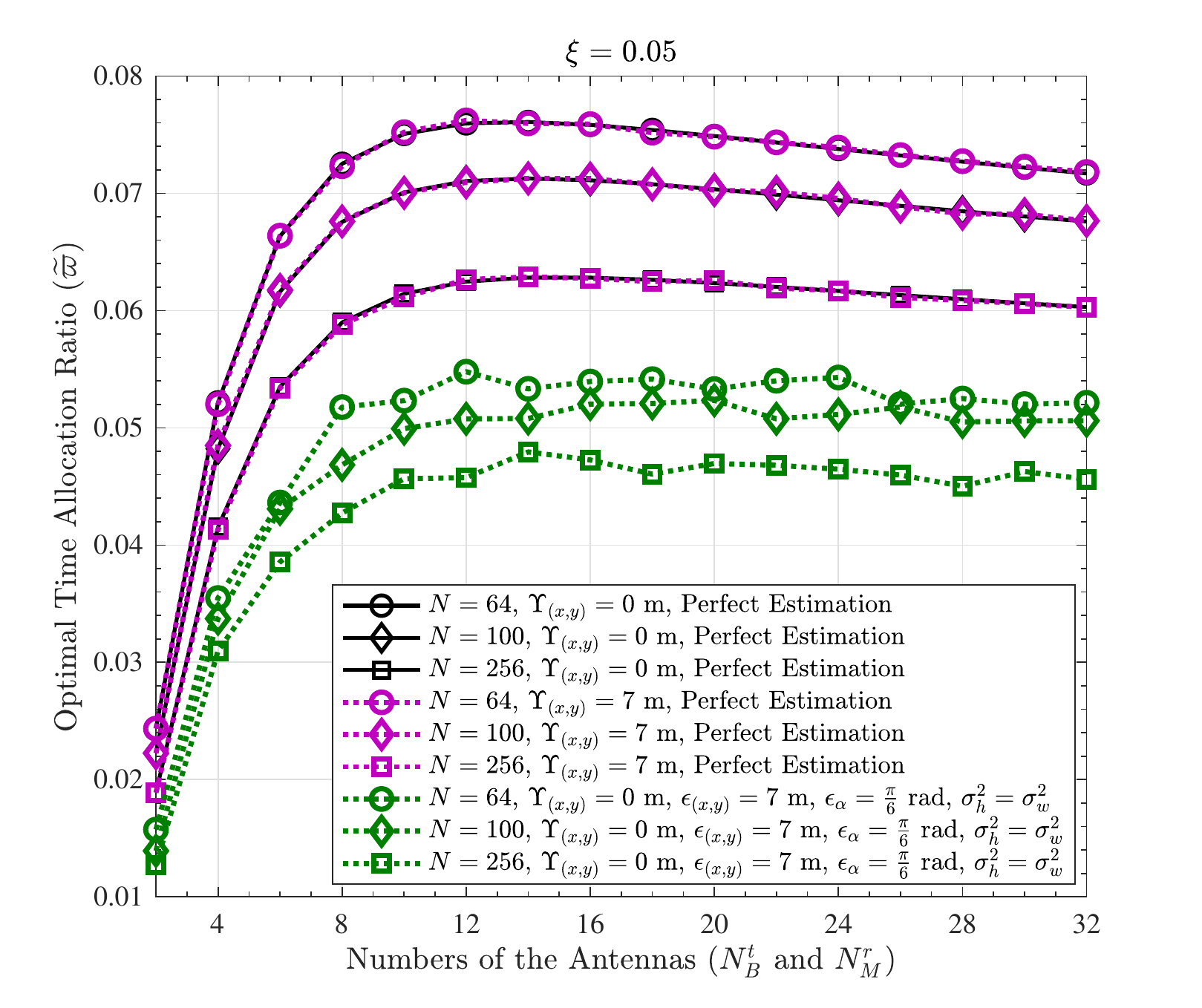}}
\label{Optimal_Varpi_AntennaNum}
\hfil
\caption{The optimal time allocation ratios ($\widetilde{\varpi}$) as functions of (a) $\sqrt{N}$, and (b) $N_B^t$ or $N_M^r$, with different $\epsilon_{(x,y)}$, $\epsilon_{\alpha}$, $\Upsilon_{(x,y)}$ and $\sigma_h^2$. The dotted curves represent the results averaged on 20 Monte Carlo trials.}
\label{Optimal_Varpi}
\end{figure*}

Then, we investigate the joint optimization performance of our proposed algorithm in Section IV. Figure \ref{Optimal_Varpi} depicts the optimal time allocation ratio ($\widetilde{\varpi}$) by varying $\sqrt{N}$ in Figure \ref{Optimal_Varpi} (a) and $N_B^t$ or $N_M^r$ in Figure \ref{Optimal_Varpi} (b), in the presence of different levels of estimation errors and user mobility. Figure \ref{Optimal_Varpi} (a) indicates that as $\sqrt{N}$ grows, $\widetilde{\varpi}$ decreases, demonstrating that more time should be allocated for the EDTS. Figure \ref{Optimal_Varpi} (b) indicates that as $N_B^t$ or $N_M^r$ grows, $\widetilde{\varpi}$ first increases rapidly and then decreases slowly, hinting that when the BS and MU are equipped with less than 8 antennas, adding more antennas will significantly alter $\widetilde{\varpi}$ by allocating more time for the BALS. Both Figure \ref{Optimal_Varpi} (a) and (b) reveal that the user mobility with $\Upsilon_{(x,y)}\leq 7$ m hardly influences $\widetilde{\varpi}$, while the estimation errors of the position/orientation and channel coefficient reduce $\widetilde{\varpi}$ to some extent.

\begin{figure}[!t]
\includegraphics[width=3.2in]{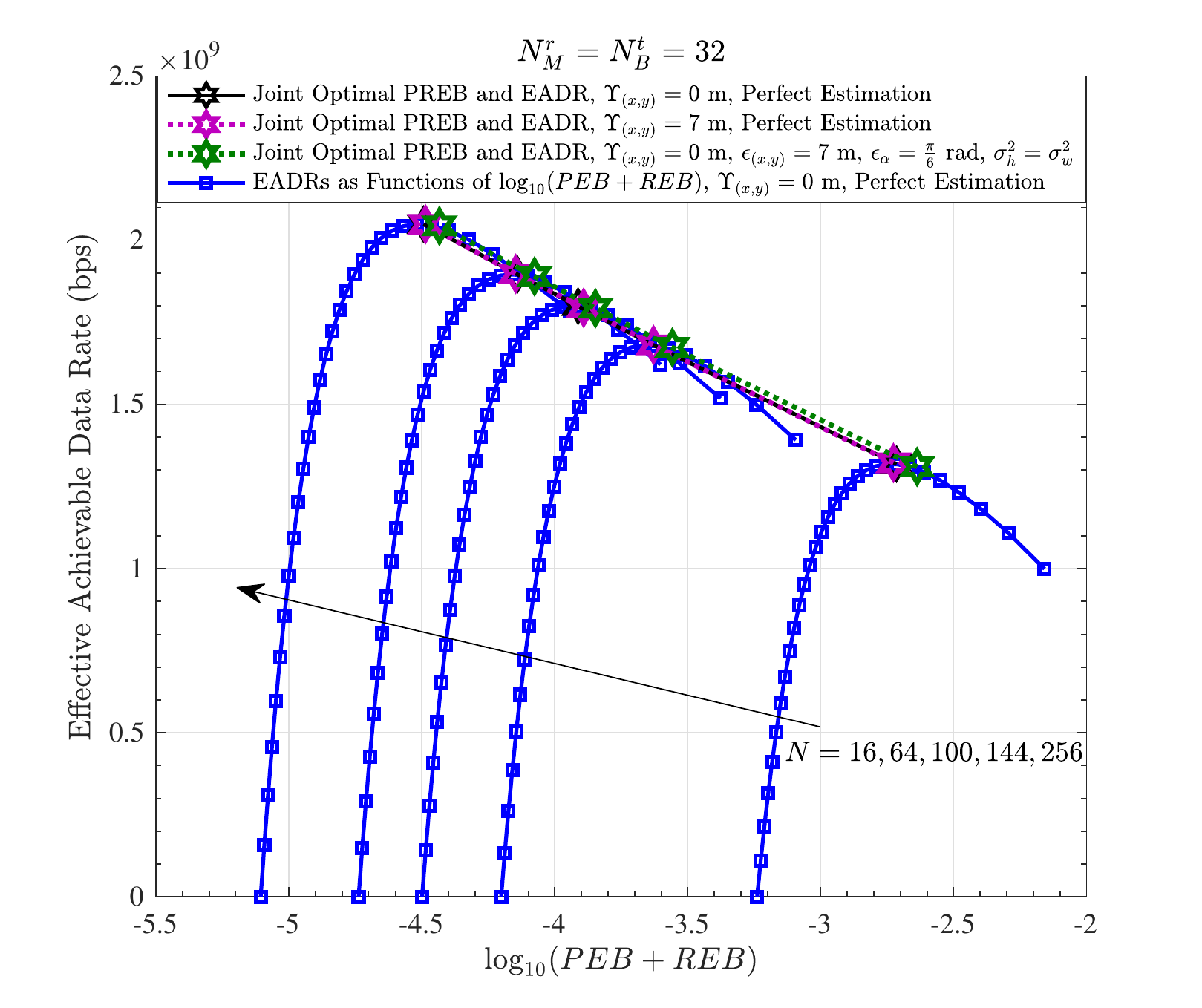}
\hfil
\centering
\caption{Joint optimal EADRs and $\log_{10}(PEB+REB)$ with different $\epsilon_{(x,y)}$, $\epsilon_{\alpha}$, $\Upsilon_{(x,y)}$ and $\sigma_h^2$. The dotted curves represent the results averaged on 20 Monte Carlo trials.}
\label{Optimal_PREB_EADR}
\end{figure}

Figure \ref{Optimal_PREB_EADR} depicts the joint optimal EADRs and $\log_{10}(PEB+REB)$, which are obtained by substituting $\widetilde{\varpi}$ into (\ref{PEB_Closedform}), (\ref{REB_Closedform}) and (\ref{EADR_max}) after running \textbf{Algorithm 1}. It is shown that under the assumption of perfect estimation and $\Upsilon_{(x,y)}=0$ m, the joint optimal EADRs and $\log_{10}(PEB+REB)$ are on the peaks of the blue curves, which stand for the EADRs as functions of $\log_{10}(PEB+REB)$. This validates the effectiveness of our proposed algorithm. Moreover, the joint optimal EADRs and $\log_{10}(PEB+REB)$ with $\Upsilon_{(x,y)}=7$ m or with $\epsilon_{(x,y)}=7$ m, $\epsilon_{\alpha}=\frac{\pi}{6}$ and $\sigma_h^2=\sigma_w^2$, are close to those with $\Upsilon_{(x,y)}=0$ m and perfect estimation, demonstrating that our proposed algorithm is insensitive to slight estimation errors and user mobility.

\section{Conclusion}

In this article, by adopting the IRS to assist the mmWave-MIMO based wireless communication and localization in the 3D scenario, we first built an IMM-JLCS and designed its working process, then derived the approximate closed-form expressions of its PEB/REB and EADR with respect to the time allocation ratio of the BALS, subsequently investigated the trade-off between the two performance metrics, and finally proposed a joint optimization algorithm based on the Lagrangian multiplier and KKT conditions, to find the joint optimal PEB/REB and EADR, as well as the corresponding optimal time allocation ratio. The simulation results validated the effectiveness of the proposed algorithm, and its robustness to slight localization or channel estimation errors and user mobility. Consequently, the system and algorithm presented in our work would be promising in contributing to the development of the future integrated communication and localization framework.

\appendices

\begin{spacing}{1.23}

\section{The Elements in $\mathbf{J}_{\bm{\eta}}(\mathbf{w}_B,\widetilde{\mathbf{\Theta}}_1,\mathbf{w}_M)$}
In Appendix A, we provide the derivations and expressions of $J_{\eta_i,\eta_j}$ in the FIM for channel parameters. It is remarkable that according to (\ref{J_eta3}), we should first calculate the partial derivative of $u_0(t)$ on each parameter in $\bm{\eta}$, and obtain
\begin{small}
\begin{equation}
\frac{\partial u_0(t)}{\partial\tau_1}=-\sqrt{P_{TX}}\widetilde{h}_1\gamma_{RX,1}\gamma_{IRS}\gamma_{TX,1}\frac{\partial x_0(t-\tau_1)}{\partial\tau_1}
\end{equation}
\begin{equation}
\frac{\partial u_0(t)}{\partial\varphi_{RX,1}}=\sqrt{P_{TX}}\widetilde{h}_1\overline{\gamma_{RX,1}}\gamma_{IRS}\gamma_{TX,1}x_0(t-\tau_1)
\end{equation}
\begin{equation}
\frac{\partial u_0(t)}{\partial\varphi_{IRS,2}^a}=\sqrt{P_{TX}}\widetilde{h}_1\gamma_{RX,1}\overline{\gamma_{IRS,a}}\gamma_{TX,1}x_0(t-\tau_1)
\end{equation}
\begin{equation}
\frac{\partial u_0(t)}{\partial\varphi_{IRS,2}^e}=\sqrt{P_{TX}}\widetilde{h}_1\gamma_{RX,1}\overline{\gamma_{IRS,e}}\gamma_{TX,1}x_0(t-\tau_1)
\end{equation}
\begin{equation}
\frac{\partial u_0(t)}{\partial\widetilde{h}_{\mathfrak{R},1}}=\sqrt{P_{TX}}\gamma_{RX,1}\gamma_{IRS}\gamma_{TX,1}x_0(t-\tau_1)
\end{equation}
\begin{equation}
\frac{\partial u_0(t)}{\partial\widetilde{h}_{\mathfrak{I},1}}=j\sqrt{P_{TX}}\gamma_{RX,1}\gamma_{IRS}\gamma_{TX,1}x_0(t-\tau_1)
\end{equation}
\end{small}
where 
\begin{small}
\begin{equation}\label{gammaRX1}
\gamma_{RX,1}=\mathbf{w}_M^H\mathbf{a}_{RX}(\varphi_{RX,1})
\end{equation}
\begin{equation}\label{gammaTX1}
\gamma_{TX,1}=\mathbf{a}_{TX}^H(\varphi_{TX,1})\mathbf{w}_B
\end{equation}
\begin{equation}\label{gammaIRS}
\gamma_{IRS}=\mathbf{a}_{IRS}^H(\varphi_{IRS,2}^a,\varphi_{IRS,2}^e)\widetilde{\mathbf{\Theta}}_1\mathbf{a}_{IRS}(\varphi_{IRS,1}^a,\varphi_{IRS,1}^e)
\end{equation}
\begin{equation}\label{overline_gammaRX1}
\overline{\gamma_{RX,1}}=\mathbf{w}_M^H diag(\mathbf{c}_{RX,1})\mathbf{a}_{RX}(\varphi_{RX,1})
\end{equation}
\begin{equation}\label{gammaIRSa}
\begin{split}
\overline{\gamma_{IRS,a}}=\mathbf{a}_{IRS}^H(\varphi_{IRS,2}^a,\varphi_{IRS,2}^e)diag((\mathbf{c}_{IRS,2}^a)^H)\widetilde{\mathbf{\Theta}}_1
\mathbf{a}_{IRS}(\varphi_{IRS,1}^a,\varphi_{IRS,1}^e)
\end{split}
\end{equation}
\begin{equation}\label{gammaIRSe}
\begin{split}
\overline{\gamma_{IRS,e}}=\mathbf{a}_{IRS}^H(\varphi_{IRS,2}^a,\varphi_{IRS,2}^e)diag((\mathbf{c}_{IRS,2}^e)^H)\widetilde{\mathbf{\Theta}}_1
\mathbf{a}_{IRS}(\varphi_{IRS,1}^a,\varphi_{IRS,1}^e)
\end{split}
\end{equation}
\end{small}
$\!\!$with the $i$-th element in $\mathbf{c}_{RX,1}\in\mathbb{C}^{N_M}$ and the $[p+(q-1)L]$-th elements in $\mathbf{c}_{IRS,2}^a\in\mathbb{C}^{N}$ and $\mathbf{c}_{IRS,2}^e\in\mathbb{C}^{N}$ given by
\begin{small}
\begin{equation}
[\mathbf{c}_{RX,1}]_i=j\frac{2\pi d}{\lambda}(i-1)cos\varphi_{RX,1}
\end{equation}
\begin{equation}
[\mathbf{c}_{IRS,2}^a]_{p+(q-1)L}=j\frac{2\pi d}{\lambda}(p-1)cos\varphi_{IRS,2}^a sin\varphi_{IRS,2}^e
\end{equation}
\begin{equation}
\begin{split}
[\mathbf{c}_{IRS,2}^e]_{p+(q-1)L}=j\frac{2\pi d}{\lambda}[(p-1)
sin\varphi_{IRS,2}^a cos\varphi_{IRS,2}^e-(q-1)sin\varphi_{IRS,2}^e]
\end{split}
\end{equation}
\end{small}
$\!\!$where $p=1,2,...,L$ and $q=1,2,...,L$.

Due to the integral form in (\ref{J_eta3}), we should then calculate $\int_0^{T_s}x_0^*(t-\tau_1)x_0(t-\tau_1)dt$, $\int_0^{T_s}\frac{\partial x_0^*(t-\tau_1)}{\partial\tau_1}x_0(t-\tau_1)dt$ and $\int_0^{T_s}\frac{\partial x_0^*(t-\tau_1)}{\partial\tau_1}\frac{\partial x_0(t-\tau_1)}{\partial\tau_1}dt$. Thanks to the Parseval's theorem, we have
\begin{small}
\begin{equation}\label{int1}
\begin{split}
\int_0^{T_s}\!\!x_0^*(t-\tau_1)x_0(t-\tau_1)dt=\int_{-\pi B}^{\pi B}|X_0(\omega)|^2d\omega=
T_s
\end{split}
\end{equation}
\begin{equation}\label{int2}
\begin{split}
\int_0^{T_s}\frac{\partial x_0^*(t-\tau_1)}{\partial\tau_1}x_0(t-\tau_1)dt=\int_{-\pi B}^{\pi B}\omega|X_0(\omega)|^2d\omega=0
\end{split}
\end{equation}
\begin{equation}\label{int3}
\begin{split}
\int_0^{T_s}\frac{\partial x_0^*(t-\tau_1)}{\partial\tau_1}\frac{\partial x_0(t-\tau_1)}{\partial\tau_1}dt=\!\!\int_{-\pi B}^{\pi B}\!\!\omega^2|X_0(\omega)|^2d\omega
=\frac{T_s}{2\pi B}\int_{-\pi B}^{\pi B}\omega^2d\omega=\frac{\pi^2B^2}{3}T_s
\end{split}
\end{equation}
\end{small}

As a result, the 6 diagonal elements and the 15 upper triangular elements among $6\times6=36$ entries in $\mathbf{J}_{\bm{\eta}}(\mathbf{w}_B,\widetilde{\mathbf{\Theta}}_1,\mathbf{w}_M)$ are derived as
\begin{small}
\begin{equation}\label{2-J_tau1_tau1}
J_{\tau_1,\tau_1}=\frac{P_{TX}T_s\pi^2 B^2}{3N_0}|\widetilde{h}_1|^2|\gamma_{RX,1}|^2|\gamma_{IRS}|^2|\gamma_{TX,1}|^2
\end{equation}
\begin{equation}\label{5-J_Phi_RX1_Phi_RX1}
J_{\varphi_{RX,1},\varphi_{RX,1}}=\frac{P_{TX}T_s}{N_0}|\widetilde{h}_1|^2|\overline{\gamma_{RX,1}}|^2|\gamma_{IRS}|^2|\gamma_{TX,1}|^2
\end{equation}
\begin{equation}\label{6-J_Phi_IRS2_a_Phi_IRS2_a}
J_{\varphi_{IRS,2}^a,\varphi_{IRS,2}^a}=\frac{P_{TX}T_s}{N_0}|\widetilde{h}_1|^2|\gamma_{RX,1}|^2|\overline{\gamma_{IRS,a}}|^2|\gamma_{TX,1}|^2
\end{equation}
\begin{equation}\label{7-J_Phi_IRS2_e_Phi_IRS2_e}
J_{\varphi_{IRS,2}^e,\varphi_{IRS,2}^e}=\frac{P_{TX}T_s}{N_0}|\widetilde{h}_1|^2|\gamma_{RX,1}|^2|\overline{\gamma_{IRS,e}}|^2|\gamma_{TX,1}|^2
\end{equation}
\begin{equation}\label{10-J_hR1_hR1}
J_{\widetilde{h}_{\mathfrak{R},1},\widetilde{h}_{\mathfrak{R},1}}=\frac{P_{TX}T_s}{N_0}|\gamma_{RX,1}|^2|\gamma_{IRS}|^2|\gamma_{TX,1}|^2
\end{equation}
\begin{equation}\label{11-J_hI1_hI1}
J_{\widetilde{h}_{\mathfrak{I},1},\widetilde{h}_{\mathfrak{I},1}}=-\frac{P_{TX}T_s}{N_0}|\gamma_{RX,1}|^2|\gamma_{IRS}|^2|\gamma_{TX,1}|^2
\end{equation}
\begin{equation}\label{24-J_tau1_Phi_RX1}
\begin{split}
J_{\tau_1,\varphi_{RX,1}}=J_{\tau_1,\varphi_{IRS,2}^a}=J_{\tau_1,\varphi_{IRS,2}^e}=J_{\tau_1,\widetilde{h}_{\mathfrak{R},1}}=J_{\tau_1,\widetilde{h}_{\mathfrak{I},1}}=0
\end{split}
\end{equation}
%\begin{equation}\label{25-J_tau1_Phi_IRS2_a}
%\begin{split}
%J_{\tau_1,\varphi_{IRS,2}^a}=0
%\end{split}
%\end{equation}
%\begin{equation}\label{26-J_tau1_Phi_IRS2_e}
%\begin{split}
%J_{\tau_1,\varphi_{IRS,2}^e}=0
%\end{split}
%\end{equation}
%\begin{equation}\label{29-J_tau1_hR1}
%J_{\tau_1,\widetilde{h}_{\mathfrak{R},1}}=0
%\end{equation}
%\begin{equation}\label{30-J_tau1_hI1}
%J_{\tau_1,\widetilde{h}_{\mathfrak{I},1}}=0
%\end{equation}
\begin{equation}\label{46-J_Phi_RX1_Phi_IRS2_a}
\begin{split}
J_{\varphi_{RX,1},\varphi_{IRS,2}^a}=\frac{P_{TX}T_s|\widetilde{h}_{1}|^2}{N_0}\mathfrak{Re}\{\overline{\gamma_{RX,1}}^*\gamma_{IRS}^*\gamma_{TX,1}^*
\gamma_{RX,1}\overline{\gamma_{IRS,a}}\gamma_{TX,1}\}
\end{split}
\end{equation}
\begin{equation}\label{47-J_Phi_RX1_Phi_IRS2_e}
\begin{split}
J_{\varphi_{RX,1},\varphi_{IRS,2}^e}=\frac{P_{TX}T_s|\widetilde{h}_{1}|^2}{N_0}\mathfrak{Re}\{\overline{\gamma_{RX,1}}^*\gamma_{IRS}^*\gamma_{TX,1}^*
\gamma_{RX,1}\overline{\gamma_{IRS,e}}\gamma_{TX,1}\}
\end{split}
\end{equation}
\begin{equation}\label{50-J_Phi_RX1_hR1}
\begin{split}
J_{\varphi_{RX,1},\widetilde{h}_{\mathfrak{R},1}}=\frac{P_{TX}T_s}{N_0}\mathfrak{Re}\{\widetilde{h}_{1}^*\overline{\gamma_{RX,1}}^*\gamma_{IRS}^*\gamma_{TX,1}^*
\gamma_{RX,1}\gamma_{IRS}\gamma_{TX,1}\}
\end{split}
\end{equation}
\begin{equation}\label{51-J_Phi_RX1_hI1}
\begin{split}
J_{\varphi_{RX,1},\widetilde{h}_{\mathfrak{I},1}}=\frac{P_{TX}T_s}{N_0}\mathfrak{Re}\{j\widetilde{h}_{1}^*\overline{\gamma_{RX,1}}^*\gamma_{IRS}^*\gamma_{TX,1}^*
\gamma_{RX,1}\gamma_{IRS}\gamma_{TX,1}\}
\end{split}
\end{equation}
\begin{equation}\label{52-J_Phi_IRS2_a_Phi_IRS2_e}
\begin{split}
J_{\varphi_{IRS,2}^a,\varphi_{IRS,2}^e}=\frac{P_{TX}T_s|\widetilde{h}_{1}|^2}{N_0}\mathfrak{Re}\{\gamma_{RX,1}^*\overline{\gamma_{IRS,a}}^*
\gamma_{TX,1}^*\gamma_{RX,1}\overline{\gamma_{IRS,e}}\gamma_{TX,1}\}
\end{split}
\end{equation}
\begin{equation}\label{55-J_Phi_Phi_IRS2_a_hR1}
\begin{split}
J_{\varphi_{IRS,2}^a,\widetilde{h}_{\mathfrak{R},1}}=\frac{P_{TX}T_s}{N_0}\mathfrak{Re}\{\widetilde{h}_{1}^*\gamma_{RX,1}^*\overline{\gamma_{IRS,a}}^*\gamma_{TX,1}^*
\gamma_{RX,1}\gamma_{IRS}\gamma_{TX,1}\}
\end{split}
\end{equation}
\begin{equation}\label{56-J_Phi_Phi_IRS2_a_hI1}
\begin{split}
J_{\varphi_{IRS,2}^a,\widetilde{h}_{\mathfrak{I},1}}=\frac{P_{TX}T_s}{N_0}\mathfrak{Re}\{j\widetilde{h}_{1}^*\gamma_{RX,1}^*\overline{\gamma_{IRS,a}}^*\gamma_{TX,1}^*
\gamma_{RX,1}\gamma_{IRS}\gamma_{TX,1}\}
\end{split}
\end{equation}
\begin{equation}\label{59-J_Phi_Phi_IRS2_e_hR1}
\begin{split}
J_{\varphi_{IRS,2}^e,\widetilde{h}_{\mathfrak{R},1}}=\frac{P_{TX}T_s}{N_0}\mathfrak{Re}\{\widetilde{h}_{1}^*\gamma_{RX,1}^*\overline{\gamma_{IRS,e}}^*\gamma_{TX,1}^*
\gamma_{RX,1}\gamma_{IRS}\gamma_{TX,1}\}
\end{split}
\end{equation}
\begin{equation}\label{60-J_Phi_Phi_IRS2_e_hI1}
\begin{split}
J_{\varphi_{IRS,2}^e,\widetilde{h}_{\mathfrak{I},1}}=\frac{P_{TX}T_s}{N_0}\mathfrak{Re}\{j\widetilde{h}_{1}^*\gamma_{RX,1}^*\overline{\gamma_{IRS,e}}^*\gamma_{TX,1}^*
\gamma_{RX,1}\gamma_{IRS}\gamma_{TX,1}\}
\end{split}
\end{equation}
\begin{equation}\label{66-J_hR1_hI1}
J_{\widetilde{h}_{\mathfrak{R},1},\widetilde{h}_{\mathfrak{I},1}}=0
\end{equation}
\end{small}
\ \ Finally, due to the symmetry of the FIM ($J_{\eta_i,\eta_j}=J_{\eta_j,\eta_i}$), each lower triangular element equals to its corresponding upper triangular element.

\section{The Elements in $\mathbf{T}$}

Let the $(i,j)$-th element in $\mathbf{T}$ be denoted by $T_{i,j}$, where $i=1,2,3$ and $j=1,2,...,6$. Then, we have the following relations:
\begin{small}
\begin{equation}\label{T_11}
T_{1,1}=\frac{\partial\tau_1}{\partial p_x}=\frac{p_x-v_x}{c\|\mathbf{p}-\mathbf{v}\|_2}
\end{equation}
\begin{equation}\label{T_12}
T_{1,2}=\frac{\partial\varphi_{RX,1}}{\partial p_x}=\frac{cos \alpha-\frac{(p_x-v_x)[(p_x-v_x)cos\alpha-(p_y-v_y)sin\alpha]}{\|\mathbf{p}-\mathbf{v}\|_2^2}}
{\sqrt{\|\mathbf{p}-\mathbf{v}\|_2^2-[(p_x-v_x)cos\alpha-(p_y-v_y)sin\alpha]^2}}
\end{equation}
\begin{equation}\label{T_13}
T_{1,3}=\frac{\partial\varphi_{IRS,2}^a}{\partial p_x}=
-\frac{p_y-v_y}{(p_x-v_x)^2+(p_y-v_y)^2}
\end{equation}
\begin{equation}\label{T_14}
T_{1,4}=\frac{\partial\varphi_{IRS,2}^e}{\partial p_x}=
\frac{\beta_{IRS}(p_x-v_x)}{\|\mathbf{p}-\mathbf{v}\|_2^2 \sqrt{\|\mathbf{p}-\mathbf{v}\|_2^2-\beta_{IRS}^2}}
\end{equation}
\begin{equation}\label{T_15}
\begin{split}
T_{1,5}=\frac{\partial\widetilde{h}_{\mathfrak{R},1}}{\partial p_x}=-\mathfrak{Re}(h_1)\zeta \left(\frac{\lambda}{4\pi}\right)
\left(\|\mathbf{v}-\mathbf{q}\|_2+\|\mathbf{p}-\mathbf{v}\|_2\right)^{-2} \frac{(p_x-v_x)}{\|\mathbf{p}-\mathbf{v}\|_2}
\end{split}
\end{equation}
\begin{equation}\label{T_16}
\begin{split}
T_{1,6}=\frac{\partial\widetilde{h}_{\mathfrak{I},1}}{\partial p_x}=-\mathfrak{Im}(h_1)\zeta \left(\frac{\lambda}{4\pi}\right)
\left(\|\mathbf{v}-\mathbf{q}\|_2+\|\mathbf{p}-\mathbf{v}\|_2\right)^{-2} \frac{(p_x-v_x)}{\|\mathbf{p}-\mathbf{v}\|_2}
\end{split}
\end{equation}
\begin{equation}\label{T_21}
T_{2,1}=\frac{\partial\tau_1}{\partial p_y}=\frac{p_y-v_y}{c\|\mathbf{p}-\mathbf{v}\|_2}
\end{equation}
\begin{equation}\label{T_22}
T_{2,2}=\frac{\partial\varphi_{RX,1}}{\partial p_y}=-\frac{sin \alpha+\frac{(p_y-v_y)[(p_x-v_x)cos\alpha-(p_y-v_y)sin\alpha]}{\|\mathbf{p}-\mathbf{v}\|_2^2}}
{\sqrt{\|\mathbf{p}-\mathbf{v}\|_2^2-[(p_x-v_x)cos\alpha-(p_y-v_y)sin\alpha]^2}}
\end{equation}
\begin{equation}\label{T_23}
T_{2,3}=\frac{\partial\varphi_{IRS,2}^a}{\partial p_y}=
\frac{p_x-v_x}{(p_x-v_x)^2+(p_y-v_y)^2}
\end{equation}
\begin{equation}\label{T_24}
T_{2,4}=\frac{\partial\varphi_{IRS,2}^e}{\partial p_y}=
\frac{\beta_{IRS}(p_y-v_y)}{\|\mathbf{p}-\mathbf{v}\|_2^2 \sqrt{\|\mathbf{p}-\mathbf{v}\|_2^2-\beta_{IRS}^2}}
\end{equation}
\begin{equation}\label{T_25}
\begin{split}
T_{2,5}=\frac{\partial\widetilde{h}_{\mathfrak{R},1}}{\partial p_y}=-\mathfrak{Re}(h_1)\zeta \left(\frac{\lambda}{4\pi}\right)
\left(\|\mathbf{v}-\mathbf{q}\|_2+\|\mathbf{p}-\mathbf{v}\|_2\right)^{-2} \frac{(p_y-v_y)}{\|\mathbf{p}-\mathbf{v}\|_2}
\end{split}
\end{equation}
\begin{equation}\label{T_26}
\begin{split}
T_{2,6}=\frac{\partial\widetilde{h}_{\mathfrak{I},1}}{\partial p_y}=-\mathfrak{Im}(h_1)\zeta \left(\frac{\lambda}{4\pi}\right)
\left(\|\mathbf{v}-\mathbf{q}\|_2+\|\mathbf{p}-\mathbf{v}\|_2\right)^{-2} \frac{(p_y-v_y)}{\|\mathbf{p}-\mathbf{v}\|_2}
\end{split}
\end{equation}
\begin{equation}\label{T_31}
T_{3,1}=\frac{\partial\tau_1}{\partial \alpha}=0
\end{equation}
\begin{equation}\label{T_32}
T_{3,2}=\frac{\partial\varphi_{RX,1}}{\partial \alpha}=-\frac{(p_x-v_x)sin\alpha+(p_y-v_y)cos\alpha}
{\sqrt{\|\mathbf{p}-\mathbf{v}\|_2^2-[(p_x-v_x)cos\alpha-(p_y-v_y)sin\alpha]^2}}
\end{equation}
\begin{equation}\label{T_33}
T_{3,3}=\frac{\partial\varphi_{IRS,2}^a}{\partial \alpha}=T_{3,4}=\frac{\partial\varphi_{IRS,2}^e}{\partial \alpha}=T_{3,5}=\frac{\partial\widetilde{h}_{\mathfrak{R},1}}{\partial \alpha}=T_{3,6}=\frac{\partial\widetilde{h}_{\mathfrak{I},1}}{\partial \alpha}=
0
\end{equation}
%\begin{equation}\label{T_34}
%T_{3,4}=\frac{\partial\varphi_{IRS,2}^e}{\partial \alpha}=
%0
%\end{equation}
%\begin{equation}\label{T_35}
%\begin{split}
%T_{3,5}=\frac{\partial\widetilde{h}_{\mathfrak{R},1}}{\partial \alpha}=0
%\end{split}
%\end{equation}
%\begin{equation}\label{T_36}
%\begin{split}
%T_{3,6}=\frac{\partial\widetilde{h}_{\mathfrak{I},1}}{\partial \alpha}=0
%\end{split}
%\end{equation}
\end{small}
\section{Derivations of the Elements in Matrix $\mathbf{A}$}

In Appendix C, we derive the expressions of the elements in $\mathbf{A}$. As illustrated in Section II, in the BALS, the transmit beamformers and receive combining vectors are searched column-by-column from $\bm{\mathcal{C}}_{BS}$ and $\bm{\mathcal{C}}_{MU}$. Let $\mathbf{w}_B(m_B)=\left[\bm{\mathcal{C}}_{BS}\right]_{m_B}$ and $\mathbf{w}_M(m_M)=\left[\bm{\mathcal{C}}_{MU}\right]_{m_M}$ denote the $m_B$-th searched transmit beamformer and the $m_M$-th searched receive combining vector, respectively, which are specifically given in (\ref{m_B_column}) and (\ref{m_M_column}). Based on these definitions, first, we calculate $\mathbb{E}_{(\mathbf{w}_B,\mathbf{w}_M)}\left[\gamma_{TX,1}\right]$. According to (\ref{gammaTX1}), we have
\begin{small}
\begin{equation}
\begin{split}
\mathbb{E}_{(\mathbf{w}_B,\mathbf{w}_M)}\left[\gamma_{TX,1}\right]
&=\mathbb{E}_{m_B}\left[\mathbf{a}_{TX}^H(\varphi_{TX,1})\mathbf{w}_B(m_B)\right]\\
&=\frac{1}{\sqrt{N_B}}\mathbb{E}_{m_B}\left[1+e^{-j\Delta\varphi_{TX,1}(m_B)}+e^{-j2\Delta\varphi_{TX,1}(m_B)}...+e^{-j(N_B-1)\Delta\varphi_{TX,1}(m_B)}\right]
\end{split}
\end{equation}
\end{small}
\!\!\! where $\Delta\varphi_{TX,1}(m_B)=\frac{2\pi}{N_B}(m_B-1)+\frac{2\pi d}{\lambda}sin\varphi_{TX,1}$. It is notable that for each $\mathbb{E}_{m_B}\left[e^{-jk\Delta\varphi_{TX,1}(m_B)}\right]$ for $k=1,2,...,N_B-1$, we have
\begin{small}
\begin{equation}
\begin{split}
\mathbb{E}_{m_B}\left[e^{-jk\Delta\varphi_{TX,1}(m_B)}\right]
&=\mathbb{E}_{m_B}\left[e^{-jk\left(\frac{2\pi}{N_B}(m_B-1)+\frac{2\pi d}{\lambda}sin\varphi_{TX,1}\right)}\right]
=\mathbb{E}_{m_B}\left[e^{-jk\frac{2\pi}{N_B}(m_B-1)}\right]e^{-jk\frac{2\pi d}{\lambda}sin\varphi_{TX,1}}\\
&=e^{-jk\frac{2\pi d}{\lambda}sin\varphi_{TX,1}}\times\frac{1}{N_B}\sum_{m_B=1}^{N_B}e^{-jk\frac{2\pi}{N_B}(m_B-1)}=0
\end{split}
\end{equation}
\end{small}
\!\!\! because $\sum_{m_B=1}^{N_B}e^{-jk\frac{2\pi}{N_B}(m_B-1)}$ is the summation performed for an entire cycle of $e^{-jk\frac{2\pi}{N_B}(m_B-1)}$. Therefore, we obtain

\begin{small}
\begin{equation}
\mathbb{E}_{(\mathbf{w}_B,\mathbf{w}_M)}\left[\gamma_{TX,1}\right]=\frac{1}{\sqrt{N_B}}
\end{equation}
\end{small}
\ \ Similarly, we also obtain
\begin{small}
\begin{equation}
\mathbb{E}_{(\mathbf{w}_B,\mathbf{w}_M)}\left[\gamma_{RX,1}\right]=\frac{1}{\sqrt{N_M}}
\end{equation}
\begin{equation}
\mathbb{E}_{(\mathbf{w}_B,\mathbf{w}_M)}\left[\overline{\gamma_{RX,1}}\right]=0
\end{equation}
\end{small}
\ \ Then, we calculate $\mathbb{E}_{(\mathbf{w}_B,\mathbf{w}_M)}\left[|\gamma_{RX,1}|^2\right]$. According to (\ref{gammaRX1}), we have
\begin{small}
\begin{equation}
\begin{split}
\mathbb{E}_{(\mathbf{w}_B,\mathbf{w}_M)}\left[|\gamma_{RX,1}|^2\right]
=&\mathbb{E}_{m_M}\left[\mathbf{w}_M^H(m_M)\mathbf{a}_{RX}(\varphi_{RX,1})\times\mathbf{w}_M^T(m_M)\mathbf{a}_{RX}^*(\varphi_{RX,1})\right]\\
=&\frac{1}{N_M}\mathbb{E}_{m_M}\left[\left(1+e^{j\Delta\varphi_{RX,1}(m_M)}+e^{j2\Delta\varphi_{RX,1}(m_M)}+...+e^{j(N_M-1)\Delta\varphi_{RX,1}(m_M)}\right)\times\right.\\
&\left.\left(1+e^{-j\Delta\varphi_{RX,1}(m_M)}+e^{-j2\Delta\varphi_{RX,1}(m_M)}+...+e^{-j(N_M-1)\Delta\varphi_{RX,1}(m_M)}\right)\right]=1
\end{split}
\end{equation}
\end{small}
where $\Delta\varphi_{RX,1}(m_M)=\frac{2\pi}{N_M}(m_M-1)+\frac{2\pi d}{\lambda}sin\varphi_{RX,1}$. Similarly, we also obtain
\begin{small}
\begin{equation}
\begin{split}
\mathbb{E}_{(\mathbf{w}_B,\mathbf{w}_M)}\left[|\gamma_{TX,1}|^2\right]
=1
\end{split}
\end{equation}
\end{small}
\ \ Subsequently, we calculate $\mathbb{E}_{(\mathbf{w}_B,\mathbf{w}_M)}\left[|\overline{\gamma_{RX,1}}|^2\right]$. According to (\ref{overline_gammaRX1}), we have
\begin{small}
\begin{equation}
\begin{split}
&\mathbb{E}_{(\mathbf{w}_B,\mathbf{w}_M)}\left[|\overline{\gamma_{RX,1}}|^2\right]=
\mathbb{E}_{m_M}\left[\mathbf{w}_M^T(m_M) diag(\mathbf{c}_{RX,1}^*)\mathbf{a}_{RX}^*(\varphi_{RX,1})\times\mathbf{w}_M^H(m_M) diag(\mathbf{c}_{RX,1})\mathbf{a}_{RX}(\varphi_{RX,1})\right]\\
=\ &\frac{1}{N_M}\mathbb{E}_{m_M}\left[\left(0+j\frac{2\pi d}{\lambda}e^{j\Delta\varphi_{RX,1}(m_M)}cos\varphi_{RX,1}+...+j\frac{2\pi d}{\lambda}(N_M-1)e^{j(N_M-1)\Delta\varphi_{RX,1}(m_M)}cos\varphi_{RX,1}\right)\times\right.\\
&\left.\left(0-j\frac{2\pi d}{\lambda}e^{-j\Delta\varphi_{RX,1}(m_M)}cos\varphi_{RX,1}-...-j\frac{2\pi d}{\lambda}(N_M-1)e^{-j(N_M-1)\Delta\varphi_{RX,1}(m_M)}cos\varphi_{RX,1}\right)\right]\\
=\ &0+\frac{1}{N_M}\left[\left(\frac{2\pi d}{\lambda}cos\varphi_{RX,1}\right)^2+\left(2\times\frac{2\pi d}{\lambda}cos\varphi_{RX,1}\right)^2+...+\left((N_M-1)\frac{2\pi d}{\lambda}cos\varphi_{RX,1}\right)^2\right]\\
=\ &\frac{1}{N_M}\times\frac{4\pi^2 d^2}{\lambda^2}[1^2+2^2+...+(N_M-1)^2]cos^2\varphi_{RX,1}=\frac{4\pi^2 d^2(N_M-1)(2N_M-1)cos^2\varphi_{RX,1}}{6\lambda^2}
\end{split}
\end{equation}
\end{small}
\ \ Finally, we calculate $\mathbb{E}_{(\mathbf{w}_B,\mathbf{w}_M)}\left[\overline{\gamma_{RX,1}}^*\gamma_{RX,1}\right]$. According to (\ref{overline_gammaRX1}) and (\ref{gammaRX1}), we have
\begin{small}
\begin{equation}
\begin{split}
&\mathbb{E}_{(\mathbf{w}_B,\mathbf{w}_M)}\left[\overline{\gamma_{RX,1}}^*\gamma_{RX,1}\right]
=\mathbb{E}_{m_M}\left[\mathbf{w}_M^T(m_M) diag(\mathbf{c}_{RX,1}^*)\mathbf{a}_{RX}^*(\varphi_{RX,1})\times\mathbf{w}_M^H(m_M) \mathbf{a}_{RX}(\varphi_{RX,1})\right]\\
=&\frac{1}{N_M}\mathbb{E}_{m_M}\left[\left(0-j\frac{2\pi d}{\lambda}e^{-j\Delta\varphi_{RX,1}(m_M)}cos\varphi_{RX,1}-...-j\frac{2\pi d}{\lambda}(N_M-1)e^{-j(N_M-1)\Delta\varphi_{RX,1}(m_M)}cos\varphi_{RX,1}\right)\times\right.\\
&\left.\left(1+e^{j\Delta\varphi_{RX,1}(m_M)}+e^{j2\Delta\varphi_{RX,1}(m_M)}+...+e^{j(N_M-1)\Delta\varphi_{RX,1}(m_M)}\right)\right]\\
=&0-\frac{1}{N_M}\left[j\frac{2\pi d}{\lambda}cos\varphi_{RX,1}+j\frac{2\pi d}{\lambda}\times 2\times cos\varphi_{RX,1}+...+j\frac{2\pi d}{\lambda}(N_M-1)cos\varphi_{RX,1}\right]\\
=&-\frac{1}{N_M}\times j\frac{2\pi d}{\lambda}\left[1+2+...+(N_M-1)\right]cos\varphi_{RX,1}
=-j\frac{\pi d(N_M-1)}{\lambda}cos\varphi_{RX,1}
\end{split}
\end{equation}
\end{small}
\ \ Therefore, we obtain the 6 diagonal elements and the 15 upper triangular elements in $\mathbf{A}$ as:
\begin{small}
\begin{equation}
A_{1,1}=\mathbb{E}_{(\mathbf{w}_B,\mathbf{w}_M)}\left[J_{\tau_1,\tau_1}\right]=\frac{P_{TX}T_s\pi^2 B^2}{3N_0}|\widetilde{h}_1|^2|\gamma_{IRS}|^2
\end{equation}
\begin{equation}
A_{2,2}=\mathbb{E}_{(\mathbf{w}_B,\mathbf{w}_M)}\left[J_{\varphi_{RX,1},\varphi_{RX,1}}\right]=\frac{4P_{TX}T_s\pi^2 d^2(N_M-1)(2N_M-1)cos^2\varphi_{RX,1}}{6\lambda^2 N_0}|\widetilde{h}_1|^2|\gamma_{IRS}|^2
\end{equation}
\begin{equation}
A_{3,3}=\mathbb{E}_{(\mathbf{w}_B,\mathbf{w}_M)}\left[J_{\varphi_{IRS,2}^a,\varphi_{IRS,2}^a}\right]=\frac{P_{TX}T_s}{N_0}|\widetilde{h}_1|^2|\overline{\gamma_{IRS,a}}|^2
\end{equation}
\begin{equation}
A_{4,4}=\mathbb{E}_{(\mathbf{w}_B,\mathbf{w}_M)}\left[J_{\varphi_{IRS,2}^e,\varphi_{IRS,2}^e}\right]=\frac{P_{TX}T_s}{N_0}|\widetilde{h}_1|^2|\overline{\gamma_{IRS,e}}|^2
\end{equation}
\begin{equation}
A_{5,5}=\mathbb{E}_{(\mathbf{w}_B,\mathbf{w}_M)}\left[J_{\widetilde{h}_{\mathfrak{R},1},\widetilde{h}_{\mathfrak{R},1}}\right]=\frac{P_{TX}T_s}{N_0}|\gamma_{IRS}|^2
\end{equation}
\begin{equation}
A_{6,6}=\mathbb{E}_{(\mathbf{w}_B,\mathbf{w}_M)}\left[J_{\widetilde{h}_{\mathfrak{I},1},\widetilde{h}_{\mathfrak{I},1}}\right]=-\frac{P_{TX}T_s}{N_0}|\gamma_{IRS}|^2
\end{equation}
\begin{equation}
\begin{split}
A_{1,j}=\mathbb{E}_{(\mathbf{w}_B,\mathbf{w}_M)}\left[J_{\tau_1,\varphi_{RX,1}}\right]&=\mathbb{E}_{(\mathbf{w}_B,\mathbf{w}_M)}\left[J_{\tau_1,\varphi_{IRS,2}^a}\right]=\mathbb{E}_{(\mathbf{w}_B,\mathbf{w}_M)}\left[J_{\tau_1,\varphi_{IRS,2}^e}\right]\\
&=\mathbb{E}_{(\mathbf{w}_B,\mathbf{w}_M)}\left[J_{\tau_1,\widetilde{h}_{\mathfrak{R},1}}\right]=\mathbb{E}_{(\mathbf{w}_B,\mathbf{w}_M)}\left[J_{\tau_1,\widetilde{h}_{\mathfrak{I},1}}\right]=0, \ \ j=2,3,...,6
\end{split}
\end{equation}
\begin{equation}
\begin{split}
A_{2,3}=\mathbb{E}_{(\mathbf{w}_B,\mathbf{w}_M)}\left[J_{\varphi_{RX,1},\varphi_{IRS,2}^a}\right]=\frac{P_{TX}T_s|\widetilde{h}_{1}|^2}{N_0}\mathfrak{Re}\{-j\frac{\pi d(N_M-1)}{\lambda}cos\varphi_{RX,1}\times\gamma_{IRS}^*
\overline{\gamma_{IRS,a}}\}
\end{split}
\end{equation}
\begin{equation}
\begin{split}
A_{2,4}=\mathbb{E}_{(\mathbf{w}_B,\mathbf{w}_M)}\left[J_{\varphi_{RX,1},\varphi_{IRS,2}^e}\right]=\frac{P_{TX}T_s|\widetilde{h}_{1}|^2}{N_0}\mathfrak{Re}\{-j\frac{\pi d(N_M-1)}{\lambda}cos\varphi_{RX,1}\times\gamma_{IRS}^*
\overline{\gamma_{IRS,e}}\}
\end{split}
\end{equation}
\begin{equation}
\begin{split}
A_{2,5}=\mathbb{E}_{(\mathbf{w}_B,\mathbf{w}_M)}\left[J_{\varphi_{RX,1},\widetilde{h}_{\mathfrak{R},1}}\right]=\frac{P_{TX}T_s}{N_0}\mathfrak{Re}\{-j\frac{\pi d(N_M-1)}{\lambda}cos\varphi_{RX,1}\times\widetilde{h}_{1}^*\}|\gamma_{IRS}|^2
\end{split}
\end{equation}
\begin{equation}
\begin{split}
A_{2,6}=\mathbb{E}_{(\mathbf{w}_B,\mathbf{w}_M)}\left[J_{\varphi_{RX,1},\widetilde{h}_{\mathfrak{I},1}}\right]=\frac{P_{TX}T_s}{N_0}\mathfrak{Re}\{\frac{\pi d(N_M-1)}{\lambda}cos\varphi_{RX,1}\times\widetilde{h}_{1}^*\}|\gamma_{IRS}|^2
\end{split}
\end{equation}
\begin{equation}
\begin{split}
A_{3,4}=\mathbb{E}_{(\mathbf{w}_B,\mathbf{w}_M)}\left[J_{\varphi_{IRS,2}^a,\varphi_{IRS,2}^e}\right]=\frac{P_{TX}T_s|\widetilde{h}_{1}|^2}{N_0}\mathfrak{Re}\{\overline{\gamma_{IRS,a}}^*
\overline{\gamma_{IRS,e}}\}
\end{split}
\end{equation}
\begin{equation}
\begin{split}
A_{3,5}=\mathbb{E}_{(\mathbf{w}_B,\mathbf{w}_M)}\left[J_{\varphi_{IRS,2}^a,\widetilde{h}_{\mathfrak{R},1}}\right]=\frac{P_{TX}T_s}{N_0}\mathfrak{Re}\{\widetilde{h}_{1}^*\overline{\gamma_{IRS,a}}^*
\gamma_{IRS}\}
\end{split}
\end{equation}
\begin{equation}
\begin{split}
A_{3,6}=\mathbb{E}_{(\mathbf{w}_B,\mathbf{w}_M)}\left[J_{\varphi_{IRS,2}^a,\widetilde{h}_{\mathfrak{I},1}}\right]=\frac{P_{TX}T_s}{N_0}\mathfrak{Re}\{j\widetilde{h}_{1}^*\overline{\gamma_{IRS,a}}^*
\gamma_{IRS}\}
\end{split}
\end{equation}
\begin{equation}
\begin{split}
A_{4,5}=\mathbb{E}_{(\mathbf{w}_B,\mathbf{w}_M)}\left[J_{\varphi_{IRS,2}^e,\widetilde{h}_{\mathfrak{R},1}}\right]=\frac{P_{TX}T_s}{N_0}\mathfrak{Re}\{\widetilde{h}_{1}^*\overline{\gamma_{IRS,e}}^*
\gamma_{IRS}\}
\end{split}
\end{equation}
\begin{equation}
\begin{split}
A_{4,6}=\mathbb{E}_{(\mathbf{w}_B,\mathbf{w}_M)}\left[J_{\varphi_{IRS,2}^e,\widetilde{h}_{\mathfrak{I},1}}\right]=\frac{P_{TX}T_s}{N_0}\mathfrak{Re}\{j\widetilde{h}_{1}^*\overline{\gamma_{IRS,e}}^*
\gamma_{IRS}\}
\end{split}
\end{equation}
\begin{equation}
A_{5,6}=\mathbb{E}_{(\mathbf{w}_B,\mathbf{w}_M)}\left[J_{\widetilde{h}_{\mathfrak{R},1},\widetilde{h}_{\mathfrak{I},1}}\right]=0
\end{equation}
\end{small}
\ \ Due to the symmetry of $\mathbf{A}$, each lower triangular element equals to its corresponding upper triangular element.

\end{spacing}

\ifCLASSOPTIONcaptionsoff
  \newpage
\fi


%\begin{IEEEbiography}{Zhe Xing}
%\end{IEEEbiography}


%\newpage



%\vfill


\end{spacing}
\end{document}